\documentclass[]{aa}
\usepackage[T1]{fontenc}
\usepackage[utf8]{inputenc}
\usepackage{lmodern}
\usepackage{newtxtext,newtxmath}
\usepackage{graphicx}
\usepackage{natbib}
\bibpunct{(}{)}{;}{a}{}{,} 
\setcitestyle{round}
\usepackage{ulem}
\usepackage{textcomp}
\usepackage{gensymb}
\usepackage{longtable}
\usepackage{threeparttable}
\usepackage{multicol}
\usepackage{multirow}
\usepackage{tabularx}
\usepackage{lipsum}
\usepackage{float}
\usepackage{todonotes}
\usepackage{amsmath}
\usepackage{etoolbox}
\usepackage{url}
\usepackage{mathtools}
\usepackage{colortbl}
\usepackage{booktabs}

\usepackage[most]{tcolorbox}
\usepackage[colorlinks=true,allcolors=blue]{hyperref}
\usepackage{xcolor}

\newcommand{\coldsim}{\textsc{ColdSIM}}

\newcommand{\CII}{\ensuremath{\mbox{[\ion{C}{ii}]}}}

\newcommand{\LCII}{$L_{\textup{\CII}}$}

\newcommand{\HI}{\ion{H}{i}}
\newcommand{\HII}{\ion{H}{ii}}

\newcolumntype{Y}{>{\centering\arraybackslash}X}
\newcolumntype{W}[1]{>{\centering\arraybackslash}p{#1}}

\begin{document}
\title{Atomic and molecular gas as traced by \CII\ emission}
\author{Benedetta Casavecchia,\inst{1}
        Umberto Maio,\inst{2,3}
        C\'eline P\'eroux,\inst{4,5}
        Benedetta Ciardi\inst{1}
    }
\institute{
    Max-Planck-Institut f\"ur Astrophysik, Karl-Schwarzschild-Str. 1, 85748 Garching b. M\"unchen, Germany
    \and
    INAF-Italian National Institute of Astrophysics, Observatory of Trieste, via G.~Tiepolo 11, 34143 Trieste, Italy
    \and
    IFPU-Institute for Fundamental Physics of the Universe, via Beirut 2, 34014 Trieste, Italy
    \and
    European Southern Observatory, Karl-Schwarzschild-Str. 2, 85748 Garching b. M\"unchen, Germany
    \and
    Aix Marseille Université, CNRS, Laboratoire d’Astrophysique de Marseille (LAM) UMR 7326, 13388 Marseille, France
   }
\offprints{Benedetta Casavecchia, e-mail address: benecasa@mpa-garching.mpg.de}
\authorrunning{Casavecchia et al.}
\titlerunning{ \CII\ emission, \HI\ and H$_2$ gas}
%
%
\abstract{
The latest ALMA and JWST observations provide new information on the birth and evolution of galaxies in the early Universe at the epoch of reionization.
Measurements at redshift $ z > 5$ of their cold-gas budget are particularly important because this budget is known to be the main fuel for star formation. 
A powerful tool for probing the physics characterising galaxies at high redshift is the \CII\ $158\,\rm \mu m$ emission line. Due to its low excitation potential, \CII\ emission can be produced in photodissociation regions, neutral atomic gas, and molecular clouds.
To properly capture the cold-gas processes taking place in these environments (molecule formation, self-shielding, dust grain catalysis, and photoelectric and cosmic-ray heating), we made use of a new set of dedicated hydrodynamic simulations (\coldsim) including time-dependent non-equilibrium chemistry, star formation, stellar evolution, metal spreading, and feedback mechanisms.
We were able to accurately track the evolution of \HI, \HII\, and H$_2$ in a cosmological context and predict the contribution of each gas phase to \CII\ luminosity.
We provide formulas that can be used to estimate the mass of molecular and atomic gas from \CII\ detections.
Furthermore, we analysed the evolution of conversion factors with galactic properties, such as stellar metallicity, star formation rate, and stellar mass. We demonstrate that \CII\ emission is dominated by \HI\ gas and that most of the \CII\ luminosity is generated in warm, dense, star-forming regions.
We conclude that although \CII\  predominantly traces atomic rather than molecular gas, the \CII\ luminosity remains a robust indicator of the H$_2$ mass.
}


\maketitle

\section{Introduction}
\label{sect:intro}

The molecular phase is mainly composed of H$_2$ and is most important for the physical processes that convert gas into stars. This is traced by the history of the star formation rate (SFR). 
Neutral hydrogen provides the essential fuel, but this fuel has to cool and transform into the molecular phase for star formation. These processes are often referred to as the baryon cycle \citep{Peroux20}.

H$_2$ detections are extremely challenging because H$_2$ lacks a permanent dipole moment and because its lowest rotational transition requires temperatures of 510~K at least for excitation \citep{Saslaw67, Hollenbach79, Lepp83, Galli98, Roussel07, Fukui10, Dobbs13, Krumholz14, Togi16}. 
Hence, direct detections of molecular hydrogen 
\cite[see e.g. recent JWST determinations by][]{Hernandez23} 
trace a small portion of the warmer molecular gas, but not the majority of the cold gas (at temperatures $T < 10^2$ K).
In this context, most estimates of molecular gas masses relied on CO rotational transitions, which enable strong and easily observable emission lines.

Through the detection of CO emission, it is possible to constrain the molecular content of large samples of galaxies \citep{Saintonge11, Bolatto13, Andreani20, Tacconi20, Boogaard23, Salvestrini24}.
Unbiased surveys for molecular gas have been conducted with the Atacama Large Millimeter/submillimeter Array (ALMA) \citep{Decarli19, Decarli20}. These observations provided an assessment of the cosmic evolution of molecular gas up to redshift $z \sim 4$ \citep{Riechers19, Decarli19, Walter22, Aravena24}. 
The Very Large Array (VLA) based CO Luminosity Density at High-z (COLDz) survey made complementary measurements at higher $z$ \citep{Riechers19}, while the IRAM 30$\,$m-based xCOLD GASS provided robust $z=0$ measurements \citep{Saintonge17} and ALMACAL low-redshift measurements over a larger volume \citep{Hamanowicz23}.

However, CO is often an unreliable tracer of H$_2$ in diffuse and low-metallicity environments \cite[see e.g.][for recent studies]{Ebagezio23}. 
In these cases, the reduced dust content allows far-ultraviolet (UV) radiation from star-forming regions to penetrate deeply into molecular clouds and to photodissociate CO molecules. This creates a layer of ionised carbon (C$^+$) around a central CO core. 
In contrast, H$_2$ molecules become self-shielded from photodissociation and can survive within C$^+$ layers and in the CO core \citep{Krumholz11, Glover12, Gnedin14}. Moreover, at higher redshift, CO transitions become progressively more challenging to observe at higher frequencies (see e.g. \citealt{Feruglio23} for a broader discussion). Moreover, converting the observed CO transitions with higher rotational quantum numbers into the reference CO(1–0) introduces an uncertainty that arises from variations in the poorly understood spectral line energy distribution across different galaxy types \citep{Carilli13, Klitsch19}.
Furthermore, the $L_{{\rm CO}}$-to-$M_{\rm H_2}$ conversion factor introduces a significant systematic uncertainty, and CO may no longer serve as a reliable tracer of H$_2$ mass in extreme environments. Specifically, this conversion factor increases as the metallicity decreases because CO photodissociates to a larger depth in each cloud. Consequently, it varies not only between objects and with redshift \citep{Bolatto13}, but also in different regions.
Alternative indicators of molecular mass are often used. \cite{Dunne22} exhaustively compared the molecular gas-mass estimates from various tracers, namely the dust continuum detection and observations of CO and CI emission lines. The authors concluded that the various indicators agree well within the uncertainties. The question remains whether some or all of these indicators trace the whole molecular gas, including the more diffuse component of the gas. While H$_2$ is typically traced by CO observations, \cite{Madden20} demonstrated that in low-metallicity galaxies, gas is not self-shielded from UV photons, and this causes the formation of a layer of \CII\ -emitting gas that may not be detectable through CO observations, but might still host H$_2$. 
The authors used a photoionisation CLOUDY modelling to emphasise that these CO-dark H$_2$ regions can be traced by \CII\ or \ion{C}{i} in principle. Their findings imply that the \CII\ $158\, \mu\rm m $ line emission is an excellent tracer of the total molecular gas mass in galaxies, and this suggests that a significant reservoir of CO-dark molecular mass might have been overlooked by previous empirical determinations. This highlights the importance of the tracer used to accurately measure molecular gas in galaxies.

The fine-structure emission line of C$^+$ at a wavelength of $\sim 158~\mu$m (\CII~158~$\mu$m) has proven to be a robust tracer of gas in the early Universe, in part because of its coverage by ALMA at $ z > 4$.
Because its excitation potential is low, \CII\ emission can be produced in gas with temperatures below $\sim10^4\,\rm K$, such as neutral atomic gas and molecular gas clouds. 
Three ALMA large programs in particular have observed selected UV-bright galaxies at $z \sim 4 - 8$. The ALPINE survey \citep{LeFevre20} has reported 118 normal star-forming galaxies at $z = 4.5 - 6$, and 20 of these galaxies and 6 others were re-observed within the CRISTAL program \citep[][Herrera-Camus et al. in prep.]{Mitsuhashi24, Ikeda24, Juno24, Telikova24} in order to spatially resolve the \CII\ line and the far-infrared dust continuum emissions. The REBELS survey \citep{Bouwens22} characterised dozens of the most luminous galaxies at $z = 6.5 - 7.7$. At these redshifts, direct detections of molecular gas are challenging, and H$_2$ mass ($M_{\rm H_2}$) estimates typically rely on empirical relations that link molecular gas to the SFR \citep{Kennicutt98, Tacconi13}, infrared luminosity \citep{Leroy08, Aravena16}, dust content \citep{Remy-Ruyer14, Genzel15, Tacconi18}, and dynamical mass \citep{Daddi10}.

All these relations are empirically well calibrated in the local Universe, but are not reliable for high-redshift systems.
For this reason, a physically motivated conversion factor, 
$\alpha_{\textup{[C II]}}$, that connects the \CII\ luminosity (\LCII) to the H$_2$ mass is extremely important for an independent estimate of the molecular gas mass at the epoch of reionisation. 
Recent determinations of $\alpha_{\textup{[C II]}}$ were provided for dwarf galaxies in the local Universe \citep{Madden20} and for star-forming galaxies up to $z = 2$ \citep{Zanella18}. 
The $\alpha_{\textup{[C II]}}$ from \cite{Zanella18} was frequently applied to $z = 4 - 8$ galaxies \citep{Bethermin20, Schaerer20, Tacconi20, Dessauges-Zavadsky20, Aravena24, Kaasinen24}, but it is unclear whether \CII\ detections can be 
solid tracers of molecular gas at these high redshifts. 
A well-calibrated method for inferring $M_{\rm H_2}$ from \LCII\ 
is crucial for current and future observations. This bright emission line is the target of new facilities, including the CONCERTO instrument \citep{CONCERTO20, Gkogkou23} on the Atacama Pathfinder Experiment telescope (APEX), and future projects, such as the proposed Atacama Large Aperture Submillimeter Telescope (AtLAST) \citep{Klaassen20}.

Additionally, the physical conditions of the gas giving rise to \LCII\ is still poorly understood. It is instrumental for a full interpretation of this emission line to establish the dependences on the star-forming processes and the density of the gas traced by \CII\ . Most importantly, \LCII\ can be produced by ionised, atomic, and molecular gas, and some observational efforts attempted to disentangle the \CII\ emission arising from different phases in the Local Universe 
\citep[e.g.][]{Croxall17, Cormier19, Tarantino21}. 
Nevertheless, it is challenging to interpret the wealth of new observational data without a deeper understanding of the dominant phase of the gas it traces.

It is rather demanding to compute the cold phases of the gas in a cosmological context in a simulation. On one hand, the physics involved is complex and has many parameters; and on the other hand, the resulting computing times are extremely long. This challenge is often referred to as "molecular cosmology". Numerical works have investigated the physical mechanisms that emit \CII\ in galaxies in the local Universe and at the epoch of reionisation. 
Three main methods were commonly adopted to model the \CII\ luminosity. \cite{Lagache18} and \cite{Popping19} implemented a semi-analytical model (SAM) based on empirical relations to determine the \CII\ luminosity in galaxies from the local Universe up to $z \sim 7$ by linking their \LCII\ to the SFR. The advantage of SAMs such as those lies in the fast calculation runtime, which enables the exploration of a large parameter space. However, the simplifications inherent to SAMs can limit the accuracy of the results.
An additional method relies on zoom-in simulations of single galaxies. These were typically used to resolve small-scale processes. For instance, \cite{Vallini15}, \cite{Pallottini17}, \cite{Katz19}, \cite{Bisbas22}, and \cite{Schimek24} investigated a few individual galaxies at $z \sim 6$  with different properties, such as virial mass, metallicity, presence or absence of mergers, and resolution. 
While zoom-in simulations provide high-resolution insights into specific galaxies, they are computationally more expensive and generally limited to a small number of cases. This prevents broad statistical conclusions.
Finally, simulations that are run in cosmological boxes, although with lower resolution, can provide statistics on much larger samples of haloes.
\cite{RamosPadilla21, RamosPadilla23}, \cite{Vizgan22H2, Vizgan22HI} and \cite{Khatri24b} focused on modelling the \CII\ emission in post-processing and the contribution from the ionised atomic and molecular gas phases in the EAGLE, SIMBA and \textsc{Marigold} simulations, respectively.

Sub-grid prescriptions are crucial for modelling the cold-gas phase in zoom-in and cosmological simulations. Simple empirical approaches that are based on gas metallicity or pressure \citep{Blitz06, Krumholz13} are often used to separate neutral hydrogen into its atomic and molecular components \citep{Lagos15, Popping19, Popping22, Szakacs22}. 
Some other works employ idealised models for line emission \citep{Olsen18} or zoom-in simulations. Although these did not directly include \CII\, , they yielded valuable insights into the properties of individual haloes \citep{Pallottini19}. Therefore, it remains one of the most critical and challenging objectives for advancing our understanding of galaxy formation and evolution at the epoch of reionisation to accurately simulate the neutral and molecular phases of cold gas within a cosmological context.

In this context, the \coldsim\ simulation suite \citep[see][from now on refered to as C24]{Maio22, Maio23, Casavecchia24} proposed a novel numerical approach to investigate the physical processes underlying the \LCII\ emission that is detected in galaxies at $z>6$ within a cosmological framework. The suite consists of a set of cosmological hydrodynamic simulations incorporating a time-dependent atomic and molecular non-equilibrium chemical network coupled with star formation, stellar evolution, feedback effects, and cooling processes. All these features enable capturing the mechanisms that cause \CII\ emission in galaxies. 
In C24 we investigated the amount of C$^+$ mass density that is produced in cosmic cold gas at different epochs, as well as the relations between \LCII, SFR, and stellar mass.

The aim of this paper is to show that there are physical relations between \LCII, the atomic hydrogen mass ($M_{\rm HI}$), and the molecular hydrogen mass ($M_{\rm H_2}$).
Moreover, we investigate the evolution of the conversion factors \LCII to $M_{\rm H_2}$ ($\alpha_{\textup{[C II]}}$) and \LCII to $M_{\rm HI}$ ($\beta_{\textup{[C II]}}$) with galactic properties, such as stellar metallicity ($Z_{\star}$), SFR, and stellar mass ($M_{\star}$).

The paper is organised as follows: In Section~\ref{sect:Method} we outline our method, and Section~\ref{sect:Results} reports our findings and relative comparisons with observational works. We discuss the impacts of our results in Section~\ref{sect:Discussion} and present our main conclusions in Section~\ref{sect:Conclusions}.

\section{Method} \label{sect:Method}

In the following, we present the numerical simulations we adopted throughout this work. We describe how we extracted the expected \CII\ signal from simulated galaxies and how it can be related to \HI\ and H$_2$ gas through the $\alpha_{\textup{[C II]}}$ and $\beta_{\textup{[C II]}}$ conversion factors.

\subsection{The \coldsim\ numerical simulations}

For our analysis, we employed the \coldsim\ High Resolution (CDM~HR) and \coldsim\ Large Box (CDM~LB) simulations described in \cite{Maio23} and C24, which are part of the \coldsim\ suite that was reported in detail by \cite{Maio22}. 
The \coldsim\ simulations have an ad hoc implementation to reproduce the formation of the first structures in the Universe. Different from other approaches that were calibrated to reproduce the properties of the local Universe, \coldsim\ simulations were performed with chemical and physical processes that are typical of the early Universe. This ensured a more appropriate modelling of the pristine gas in its atomic and molecular form, and of its collapse into stars and galaxies. 
Gravity and smoothed-particle-hydrodynamics (SPH) computations were performed with a modified version of \textsc{P-Gadget3} \citep{Springel05}, which includes a unique time-dependent non-equilibrium chemical network that resolves ionisation, dissociation, and recombination processes of the primordial gas via first-order differential equations \citep{Abel97, Yoshida03, Maio07}. In this way, we tracked the evolution of the following atomic and molecular species:  e$^{-}$, H, H$^{+}$, H$^{-}$, He, He$^{+}$, He$^{++}$, H$_{2}$, H$^{+}_{2}$ , D, D$^{+}$, HD, and HeH$^{+}$.
The chemical network was evolved self-consistently with stellar feedback processes, cooling, and heating \citep{Maio22}.
At each time step, collisionless stellar particles were formed in a stochastic way following a modified version of the \cite{Springel03} model, in which we included molecular runaway cooling \citep{Maio09}.
Each of these particles represents a simple stellar population with a Salpeter initial mass function (IMF). 
Population~III (Pop~III) stars were distributed with a Salpeter IMF, similarly to population~II (Pop~II) and population~I (Pop~I).
We followed their stellar evolution according to mass and metallicity-dependent stellar yields and lifetimes.
In particular, stellar particles produce heavy elements through asymptotic giant branch (AGB) winds and Type~II and Type~Ia supernova (SN)  explosions \citep{Tornatore07, Maio10, Maio16}, including He, C, N, O, Ne, Mg, Si, S, Ca, and Fe. Each of these elements was tracked individually and mixed through the SPH kernel smoothing. 
As detailed in \cite{Maio22}, radiative losses of metals from resonant and fine-structure transitions, 
such as the \CII\ emission at $158$ $\mu$m, were included in the chemical network. Primordial gas cools through H$_2$ formation via H$^-$, H$_2^+$ catalysis, and three-body interactions in pristine gas, and additionally, via dust catalysis in metal-enriched gas particles. The H$_2$ dust-grain catalysis was performed for different gas temperatures and metallicities, assuming a $Z$-dependent dust-to-metal ratio and an energy balance between the cosmic microwave background radiation and dust-grain emission with a power law of slope $\gamma = 2$ (in agreement with recent ALMA determinations).
The standard \cite{HaardtMadau96} UV background generated by the formation of the first structures was turned on at $z = 6$.
Photoelectric and cosmic-ray heating were implemented in the chemical network and complemented with \ion{H}{i} and H$_2$ shielding \citep{Habing68, Draine87, Bakes94, Draine96}.

The simulations were performed within the standard cosmological framework based on cold dark matter and the cosmological constant $\Lambda$. 
The present-day expansion parameter was assumed to be $H_0 = 100 h$ km s$^{-1}$ Mpc$^{-1}$ with $ h = 0.7$, while the baryon, matter, and $\Lambda$ density parameters were $\Omega_{b,0} = 0.045$, $\Omega_{m,0} = 0.274$, and $\Omega_{\Lambda,0} = 0.726$, respectively.
The adopted power spectrum had a spectral index of $0.96$ and a normalisation over an 8~$\rm Mpc\,{\it h}^{-1}$ sphere of $\sigma_8 = 0.8$.
We adopted the notation cMpc to indicate lengths in comoving megaparsecs, and  unless otherwise stated, we expressed masses, luminosities, and metallicities in solar units (M$_\odot$, L$_{\odot}$, and Z$_\odot$, respectively).

The two simulated volumes for which we report results in this paper are CDM~LB and CDM~HR. 
The former has the largest volume and is able to capture the more massive galaxies. It has a size of $50$ cMpc$\,h^{\rm -1}$ and a number of  particles of $2 \times 1000^3$ that are evenly distributed between dark matter and gas at the initial redshift $z = 100$.
In CDM~LB, dark-matter and gas particles have a mass of $m_{\rm DM} = 7.9 \times 10^6$ M$_{\odot}$ and $m_{\rm gas} = 1.6 \times 10^6$ M$_{\odot}$, respectively. 
CDM~HR instead has the highest resolution because it has a size of $10$ cMpc$\,h^{\rm -1}$ and the same number of particles, resulting in $m_{\rm DM} = 6.3 \times 10^4$ M$_{\odot}$ and  $m_{\rm gas} = 1.3 \times 10^4$ M$_{\odot}$.
We show the results at $z =$ 6, 7, 8, and 10 in order to make comparisons with ALMA \CII\ observations at the epoch of reionisation and provide predictions for higher redshifts.

\begin{figure}[]
\begin{center}
  \begin{tabular}{c}
    \includegraphics[width=0.52\textwidth]{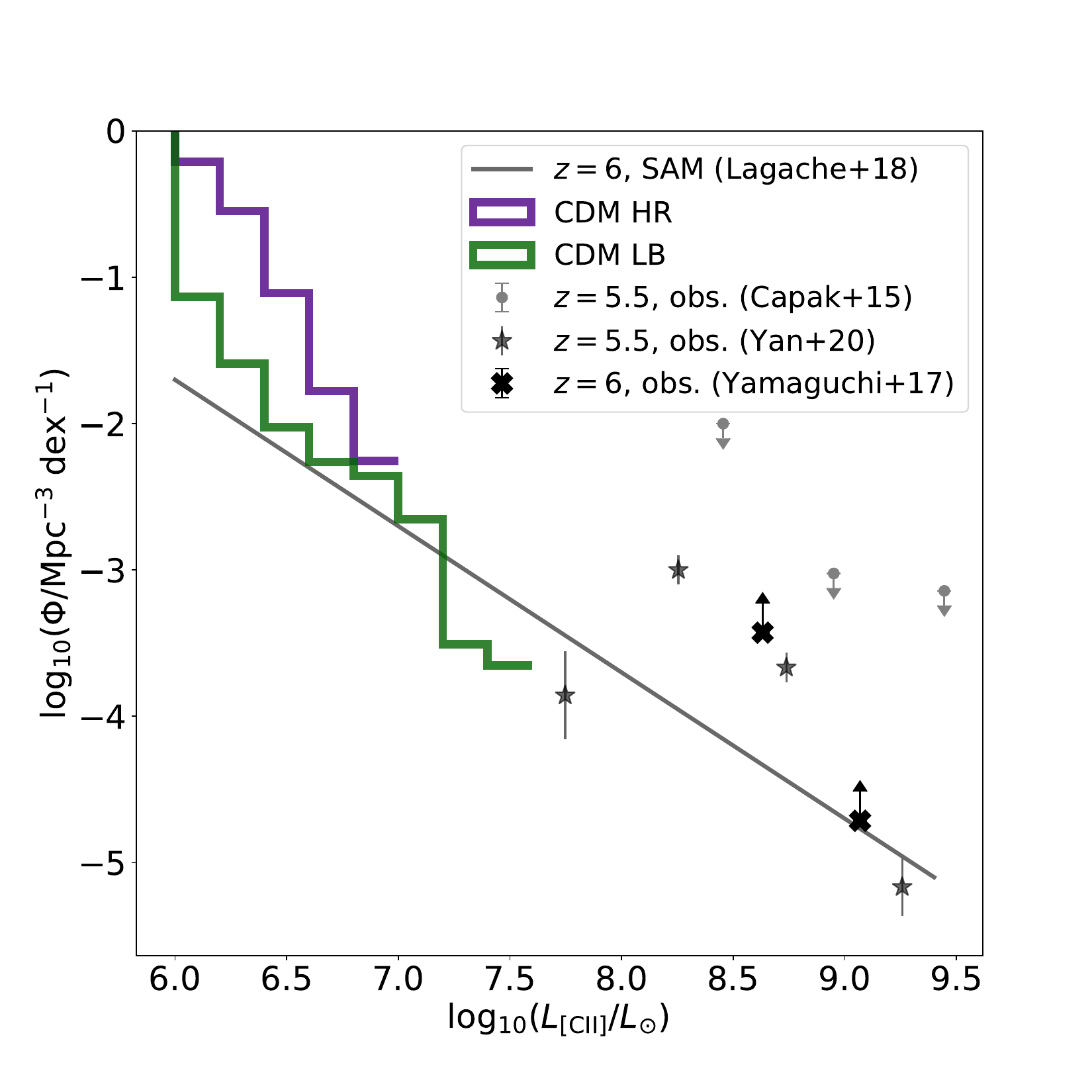}
  \end{tabular}
  \caption{\CII\ luminosity functions at $z = 6$. The coloured histograms represent predictions from CDM HR (purple) and CDM LB (green). The solid grey line shows the results from the \texttt{G.A.S.} SAM \citep{Lagache18}.
  The grey symbols show ALMA data for galaxies at $z = 5.5$ \citep[dots,][]{Capak15}, the results from the ALPINE survey \citep[stars,][]{Yan20}, and lower limits from lensed \CII\ line emitters at $z = 6$ \citep[crosses,][]{Yamaguchi17}.}
  \label{fig:Lum_funct}
\end{center}
\end{figure}

\subsection{Computing \CII\ luminosities and conversion factors}

In this section, we describe how we estimated the \CII\ emission at 158~$\mu$m for haloes in \coldsim.
The calculations for \CII\ luminosities were directly included in the simulations at runtime and contributed to cosmic gas cooling.
This guaranteed that thermodynamical variables were evaluated correctly and the \CII\ emission was derived consistently.
More detailed explanations were provided in C24, and we limit our presentation here to the most relevant methodological steps.

We considered haloes composed of at least 100 particles, and for each halo, \LCII\ was calculated as the sum of the \CII\ luminosities of each individual particle. Specifically, we modelled \CII\ as a two-level system considering the quantum number $J = 3/2$ and $J = 1/2$ states for the fine-structure transition 
$(2p)[\rm {}^2 P_{3/2} - {}^2 P_{1/2}]$.
Following for example \cite{Maio07} and C24, we can write for each gas particle
\begin{equation}
L_{\textup{\CII}} = \Lambda_{\textup{\CII}} {V} = {n}_{{\textup{C}^+,2}} \, A_{21} \, \Delta E_{21} \, {V},
\label{eq:lum}
\end{equation}
where $V$ is the volume of a sphere with a radius equal to the smoothing length of the gas particle, and $\Lambda_{\textup{\CII}}$ is the emission rate of the \CII\ fine-structure transition.
The quantity $\Lambda_{\textup{\CII}}$ is expressed as a function of the number density of C$^+$ ions excited to the upper state $n_{{\textup{C}^+,2}}$, the Einstein coefficient for spontaneous emission $A_{21}$, and the energy level separation $ \Delta E_{21} $  
(see Appendix~\ref{sect:appendix} for further details).

In Figure \ref{fig:Lum_funct} we show the \CII\ luminosity function $\Phi$ for galaxies from CDM HR and CDM LB. CDM LB has a lower mass resolution than CDM HR, but because its cosmic volume is larger, we were also able to model more massive haloes. This is clear from the higher \CII\ luminosities that were reached in this latter case ($\sim 10^{7.5} \,\rm L_\odot$). 
Because its resolution is, CDM LB was not able to fully capture the trend of the luminosity function $\Phi$ at the low-luminosity end, where the statistics of faint structures is underestimated by almost one dex in comparison to CDM HR. The faint end of \CII\ luminosity functions seems to be sensitive to the model parameters (e.g. metallicity, stellar yields, numerical resolution, and star formation stocasticity, and even to the way \CII\ calculations are performed) and to other popular diagnostics, such as UV luminosity functions \cite[][]{Maio23, Williams24, Kravtsov24, Gelli24}. The two simulations converge well around $10^7\,\rm L_\odot$.
The \texttt{G.A.S.} semi-analytical model by \cite{Lagache18} and CDM LB agree well because they rely on a similar stellar-mass resolution.
On the other hand, the CDM HR results are more robust in characterising galaxies with \LCII $<10^7~L_{\odot}$ than \texttt{G.A.S.} because the resolution of our simulation is high. Thus, CDM HR captures more galaxies with fainter \CII\ emission.

A few observational determinations at $z = 6$ can help us to constrain the \CII\ luminosity function at these redshifts \citep{Capak15, Yamaguchi17, Yan20}. These data were obtained with ALMA and cover luminosities that are slightly higher than those of \coldsim\ galaxies. 
They have $ L_{\textup{\CII}} \gtrsim 10^{7.5} \,\rm L_\odot $, and the resulting $\Phi$ features a large scatter of a few dex. In this respect, our results provide predictions for a luminosity range that is still inaccessible with current instruments. We note that \cite{Hemmati17} reported the first local \CII\ emission line luminosity function from 500 galaxies from the Revised Bright Galaxy Sample, where the \CII\ luminosities were measured from the Herschel PACS observations of luminous infrared galaxies. The authors reported an evolution in the \CII\ luminosity function with cosmic time similar to the evolution trend of the cosmic SFR density.

As discussed in the Introduction, \LCII\ is often used as tracer of atomic and molecular gas in high-$z$ galaxies. To investigate how the \CII\ emission is related to different gas phases, we introduced the \LCII-to-$M_{\rm H_2}$ conversion factor $\alpha_{\textup{[CII]}}$ and the \LCII-to-$M_{\rm HI}$ conversion factor $\beta_{\textup{[CII]}}$.
They are defined as
$\alpha_{\textup{[CII]}} = M_{\textup{H}_2}/L_{\textup{[CII]}} \quad \text{and} \quad \beta_{\textup{[CII]}} = M_{\textup{HI}}/L_{\textup{[CII]}}$, where $M_{\rm HI}$ and $M_{\rm H_2}$ are the atomic and molecular hydrogen masses of each  galaxy.
In practice, these conversion factors are mass-to-light ratios relating \CII\ to neutral gas. They are commonly reported in $ M_\odot / L_\odot $ units and allow us to link the measured \LCII\ values to the underlying H$_2$ or \HI\ gas mass.

The results from different numerical simulations in the literature do not converge yet when the molecular gas content in galaxies is estimated. In the case of \coldsim, different resolutions can lead to variations in H$_2$ masses by a factor of a few in the most extreme scenarios \citep[see][]{Maio23}.
We aim to link \CII\ emission to precise atomic and molecular mass determinations in a cosmological context. We therefore use in Section~\ref{sect:Results} the most accurate high-resolution simulation box, CDM~HR, if not otherwise stated.

\section{Results}
\label{sect:Results}

In this section, we present our results about the relations between $L_{\textup{\CII}}$ and the different gas phases, and we explore their dependence on various galactic properties, such as stellar mass, metallicity, and star formation rate.

\subsection{Relations of $L_{\textup{\CII}}$ with atomic and molecular masses }
\label{sect:linear relations}

The \CII\ luminosity is often used as tracer of cold gas (with temperatures below $10^4\,\rm K$), which is typically composed of atomic (\HI) and molecular (H$_2$) hydrogen. In this regime, \HI\ gas is expected to be dominant in mass with respect to H$_2$. However, because of the density and temperature dependences of the \CII\ emission signal (see previous section), this does not necessarily mean that most of the \LCII\ is produced by neutral atomic gas.

\begin{figure}[h]
    \centering
    \vspace{-3em}
    \includegraphics[width=0.5\textwidth]{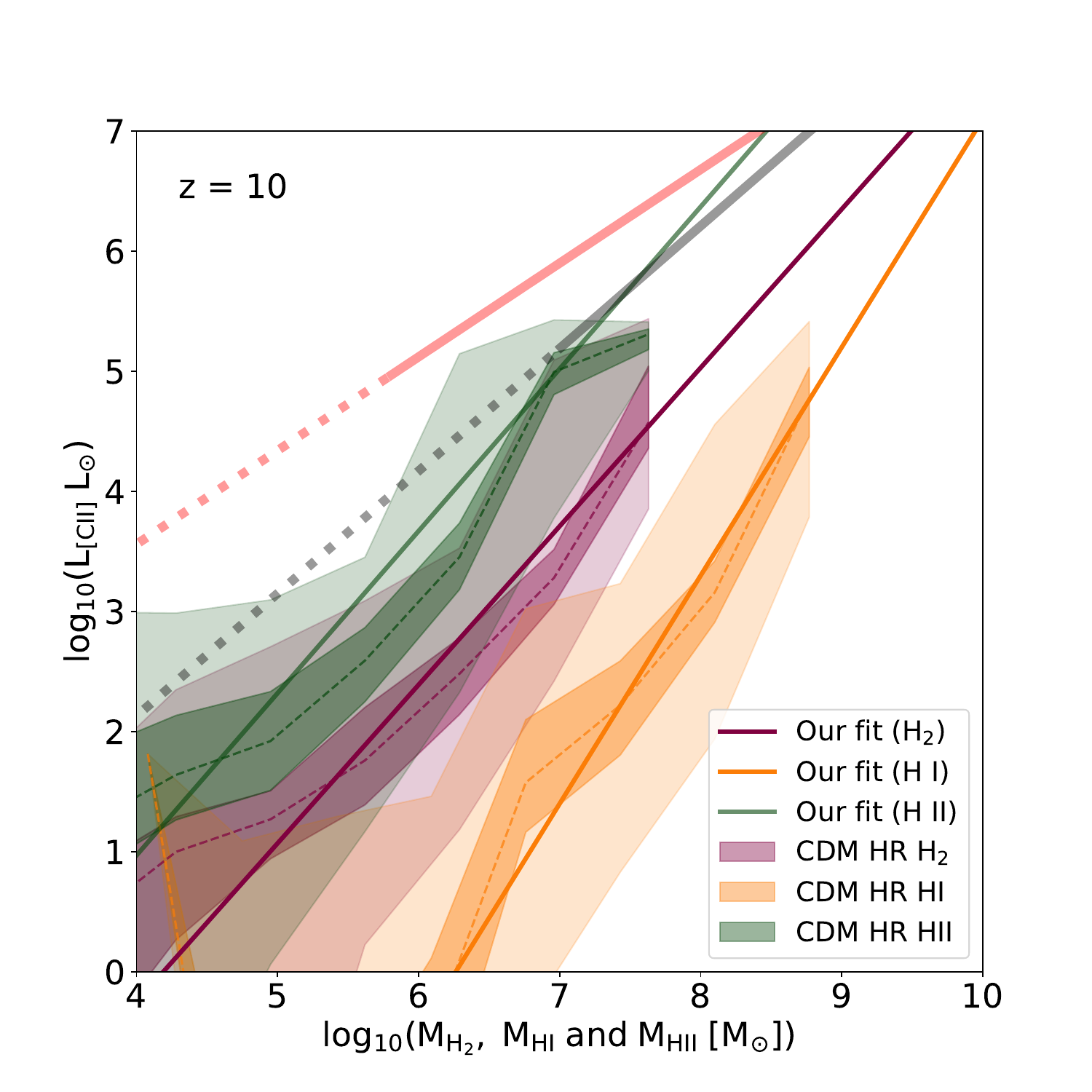}
    
    \vspace{-3em} 
    
    \includegraphics[width=0.5\textwidth]{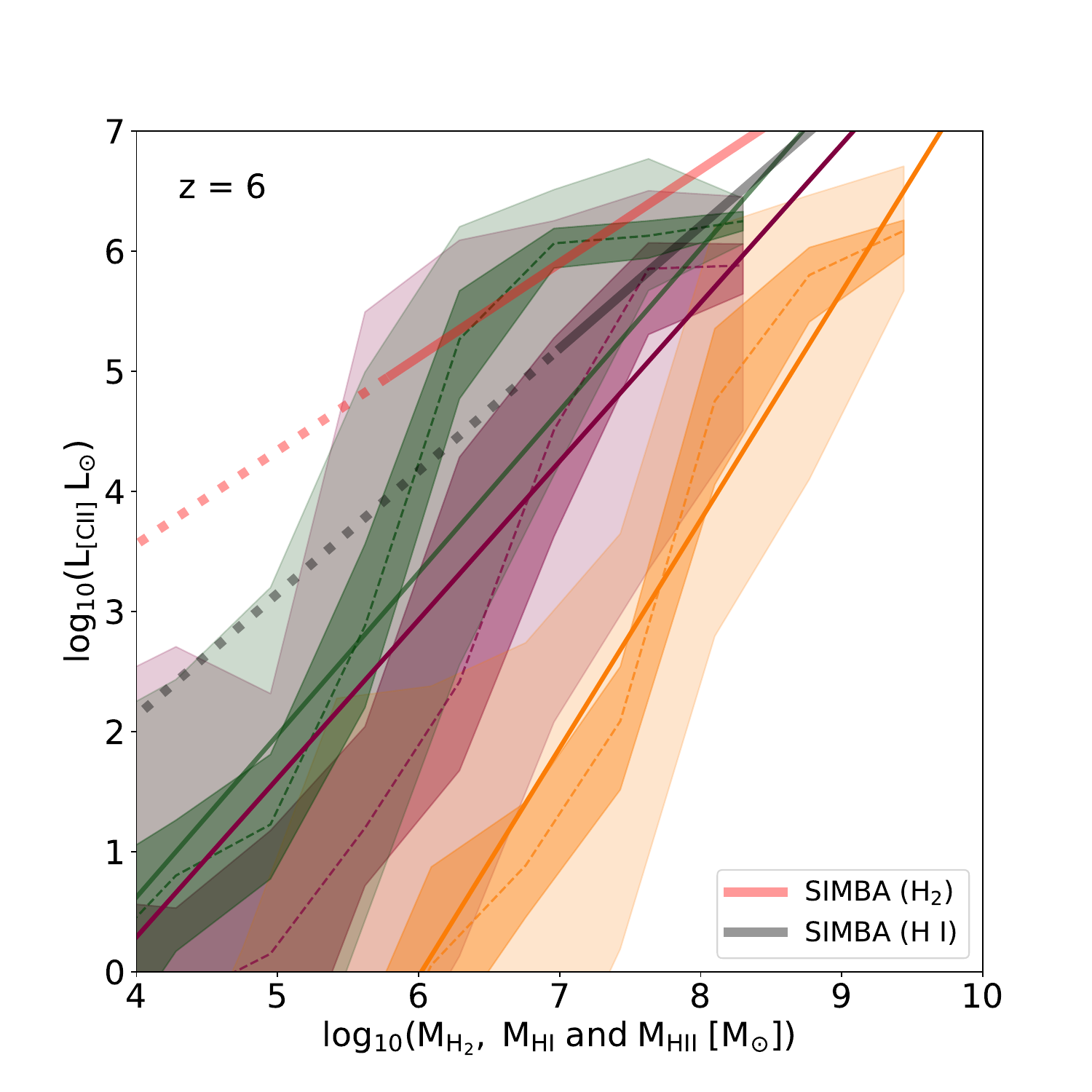}
    
    \caption{\CII\ luminosity \LCII\ as a function of H$_2$ mass $M_{\rm H_2}$ (dark pink), \HI\ mass $M_{\rm HI}$ (orange), and \HII\ mass $M_{\rm HII}$ (green) at $z = 10$ (top) and $6$ (bottom). The dashed lines show the median value for CDM HR, and the darker (lighter) shaded regions represent the 1$\sigma$ (2$\sigma$) standard deviation. The solid dark pink line represents the fit to the CDM HR results for $M_{\rm H_2}$, the solid orange line shows the fit for $M_{\rm HI}$, and the green line shows $M_{\rm HII}$ (see the text). 
    The solid red and black lines show the linear fit to post-processed SIMBA simulations for $M_{\rm H_2}$ and $M_{\rm HI}$ in their validity range \citep[see][]{Vizgan22H2, Vizgan22HI}, and the extrapolations of the fit to lower masses is highlighted with dotted lines. We highlight the redshift evolution of the linear relation between $\log_{10}(L_{\textup{\CII}})$ and  $\log_{10}(M_{\rm H_2})$, $\log_{10}(M_{\rm HI})$, and $\log_{10}(M_{\rm HII})$. Moreover, we find that \HI\ is always the dominant phase for \coldsim\ galaxies.
    }
    \label{fig:L_CIIvsMH2andMHI}
\end{figure}

In Figure \ref{fig:L_CIIvsMH2andMHI} we plot \LCII\ as a function of $M_{ \rm H_2}$, $M_{ \rm HI}$, and $M_{ \rm HII}$ at $z = 10$ and $6$. As expected, the \HII\ phase contributes least to the total mass of hydrogen in these galaxies. Concerning the neutral gas, we observe that galaxies with higher molecular and atomic masses are more luminous in \CII.
This is expected from the cooling of atomic and molecular gas and the consequent collapse into stars, which is consistent with the linear trend found in C24 between the logarithms of \LCII\ and SFR. 
A similar correlation is shown in Figure \ref{fig:L_CIIvsMH2andMHI} between \LCII\ versus $M_{ \rm H_2}$ and $M_{\rm HI}$ at both redshifts. 
The \LCII\ versus $M_{ \rm H_2}$ relation extends to higher molecular masses and \LCII\ at lower redshifts as a consequence of the galaxy build-up through gas streams and mergers. It shows a 1$\sigma$ scatter of up to $\sim$ 1.5 dex, suggesting that molecular hydrogen is more sensitive to feedback processes, and it is thus more challenging to constrain than atomic gas.
An analogous trend is also observed for the \LCII\ versus $M_{ \rm HI}$ relation.

The \coldsim\ galaxies with higher \LCII\ can retain more cold gas, atomic and molecular, and fuel the global star formation activity. 
At all redshifts and for each \LCII\ value, $M_{\rm HI}$ is always $\sim 2$~dex higher than $M_{ \rm H_2}$ 
and roughly $\sim 3$ dex higher than  $M_{\rm HII}$.
Only a small fraction (locally up to $\approx 10^{-2} - 10^{-1}$ ) of cold hydrogen gas becomes molecular, while the majority is atomic. Moreover, the scatter exhibited by the \LCII\ vs. $M_{ \rm HI}$ relation is slightly smaller than that from the \LCII\ versus $M_{ \rm H_2}$ relation.
Below, we provide linear fits, displayed in Figure \ref{fig:L_CIIvsMH2andMHI}, for the logarithms of \LCII\ versus $M_{\rm H_2}$, $M_{\rm HI}$, and $M_{\rm HII}$ gas masses that also include redshift evolution,

\begin{equation}
\begin{aligned}
    \hspace{1em}
    \log_{10}\left (L_{\textup{\CII}}/{\rm L_{\odot}}\right ) = & 1.36(\pm 0.01)~\log_{10} \left(M_{ {\rm H_2}}/{\rm M_{\odot}}\right )\\[0.5em] 
    &\hspace{1em} 
     -0.14(\pm 0.01)~z-4.13(\pm0.09)
\end{aligned}
\label{eq:H2 fit}
\end{equation}

\begin{equation}
\begin{aligned}
    \hspace{1em}
    \log_{10}\left (L_{\textup{\CII}}/{\rm L_{\odot}}\right ) = & 1.94(\pm 0.01)~\log_{10} \left(M_{ {\rm HI}}/{\rm M_{\odot}}\right )\\[0.5em] 
     &\hspace{1em}
     -0.12(\pm 0.01)~z-10.7(\pm0.1)
\end{aligned}
\label{eq:HI fit}
\end{equation}

\begin{equation}
\begin{aligned}
    \hspace{1em}
    \log_{10}\left (L_{\textup{\CII}}/{\rm L_{\odot}}\right ) = & 1.45(\pm 0.06)~\log_{10} \left(M_{ {\rm HII}}/{\rm M_{\odot}}\right )\\[0.5em] 
     &\hspace{1em}
     +0.06(\pm 0.04)~z-5.58(\pm0.43)
\end{aligned}
\label{eq:HII fit}
\end{equation}

\noindent
These fits were estimated by including all the snapshots presented in this paper ($z = $~10, 8, 7, and 6), and the uncertainty of the various parameters corresponds to one standard deviation of the residuals.
Equations \ref{eq:H2 fit}, \ref{eq:HI fit}, and \ref{eq:HII fit} can be used to estimate the mass of molecular, neutral, and ionised hydrogen from the detected \CII\ emission in high-redshift galaxies. These expressions clearly point out that \LCII\ increases with H$_2$, \HI\, and \HII\ gas masses over-linearly.
Furthermore, the redshift correction shows that for any fixed mass, earlier objects have suppressed \LCII\ by a factor of $\approx 10^{-0.1\,z}$ for H$_2$, \HI\, and \HII. This is a consequence of the typically lower metallicities that are encountered in primordial epochs and of their evolution towards lower $z$.

Our formulae for molecular and atomic hydrogen can be directly compared to literature studies, such as the relations derived from the SIMBA simulations at $z = 6$ \citep{Vizgan22H2, Vizgan22HI}. 
The authors combined the output of simulations of different box sizes to cover a wider mass range, and they estimated \LCII, $M_{ \rm H_2}$, and $M_{ \rm H I}$ in post-processing. Their fits were computed for galaxies with a molecular hydrogen mass $M_{\rm H_2} > 10^6$ M$_{\odot}$ and an atomic hydrogen mass $M_{\textup{H I}} > 10^7$~M$_{\odot}$.
In the figure, we show their results and extrapolate them to lower H$_{2}$ and \HI\ masses.
\coldsim\ galaxies are dominated in mass by atomic gas, and the \CII\ emission is mostly due to collisions with atomic hydrogen and not molecular hydrogen. 
In \cite{Vizgan22H2, Vizgan22HI}, however, molecular gas is the dominant phase, and the majority of \CII\ emission is expected to originate by interactions with H$_2$. The authors recognised that in their model, atomic gas does not contribute {\it by construction} to \LCII, however. Some of the \HI\ gas might be hidden within the ionised gas, so that the true neutral atomic mass is likely higher than indicated by their results.

\begin{figure}[]
\begin{center}
  \begin{tabular}{c}
  \includegraphics[width=0.482\textwidth]{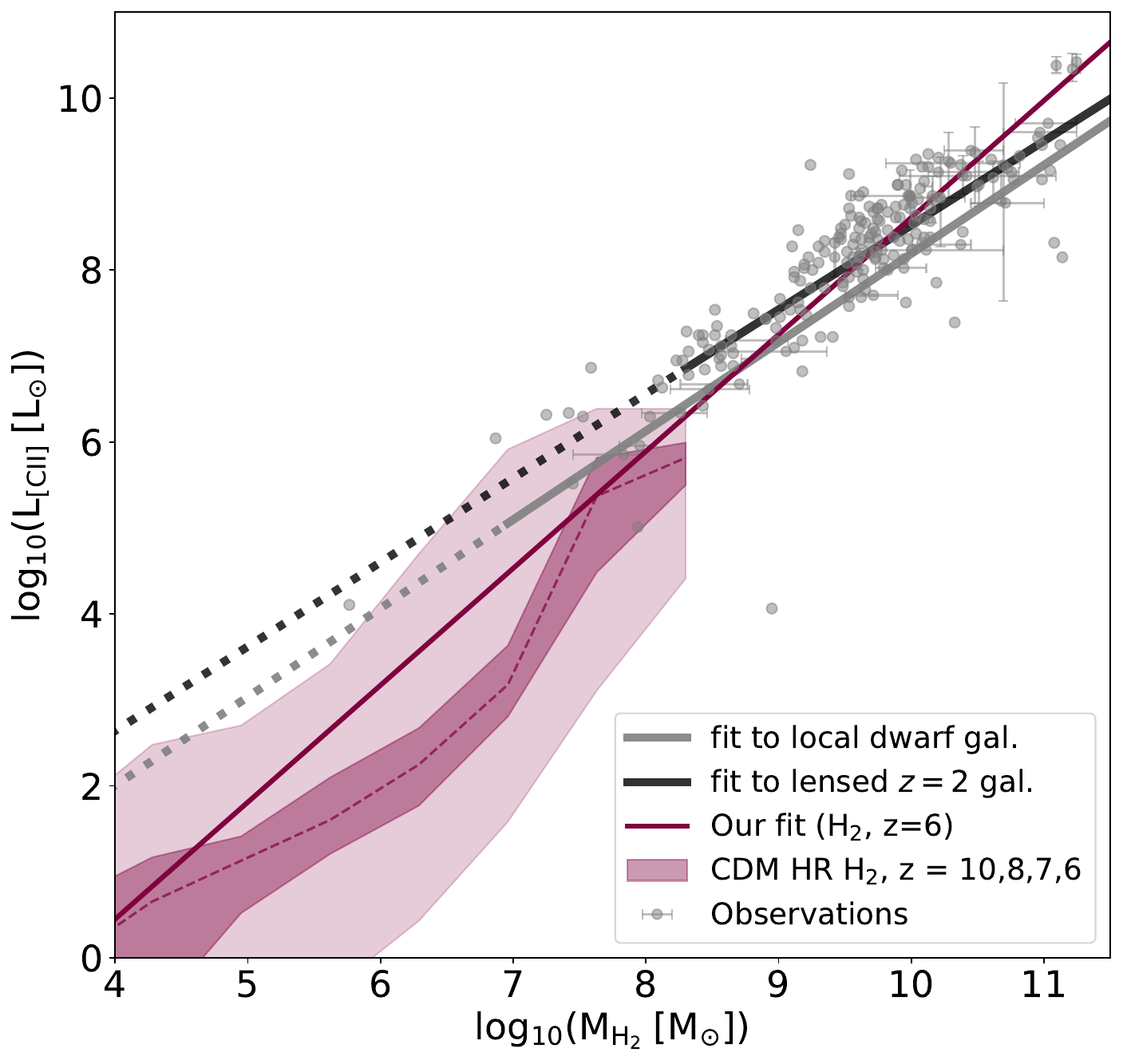}
  \end{tabular}
  \caption{\CII\ luminosity \LCII\ as a function of H$_2$ mass $M_{\rm H_2}$ at $z = 10,8,7$, and 6 combined. The dashed line refers to the mean value for CDM HR (dark pink), and the darker (lighter) shaded region represents the 1$\sigma$ (2$\sigma$) standard deviation. The solid dark pink line represents the linear fit to the CDM HR results at $z = 6$. The solid black and grey lines refer to the fits to observations by \cite{Zanella18} and \cite{Madden20} ,and the extrapolations of the fits to lower luminosities are highlighted with dotted lines. The grey dots represent observations of local dwarf galaxies \citep{Cormier15, Madden20}, $z = 0.5$ main-sequence and starburst galaxies \citep{Stacey91, DiazSantos13, DiazSantos17, Magdis14, Accurso17, Contursi17, Hughes17}, and star-forming galaxies in the range $z = 2-6$ \citep{Ferkinhoff14, Huynh14, Capak15, Gullberg15, Schaerer15, Zanella18, Kaasinen24}. Globally, we find that our simulations agree well with observations.}
  \label{fig:LCIIvsMH2_z6}
\end{center}
\end{figure}

In Figure \ref{fig:LCIIvsMH2_z6} we show the relation between \LCII\ and $M_{\rm H_2}$ of $z = 10, 8 ,7$, and 6 combined, and we compare our results with observational determinations. 
We report a correlation between \LCII\ and $M_{\rm H_2}$ that agrees well with a compilation of observations from the literature \citep{Stacey91, DiazSantos13, DiazSantos17, Ferkinhoff14, Huynh14, Magdis14, Capak15, Cormier15, Gullberg15, Schaerer15, Accurso17, Contursi17, Hughes17, Zanella18, Madden20, Kaasinen24}.
These observations included different types of galaxies: local dwarves, and main-sequence and star-forming galaxies that covered a wide range of redshift from $z = 0$ to $7$. Although these observations covered molecular gas masses and \LCII\ slightly higher than those of the \coldsim\ simulations, our results are consistent with the data of local dwarf galaxies from \cite{Cormier15} and  \cite{Madden20} and with the data of main-sequence galaxies from \cite{Accurso17}. 
The extrapolation to higher masses of our fit in equation \ref{eq:H2 fit} agrees with most observations.
In addition, we compared it to the empirical fits by \cite{Zanella18} and the fit to a model by \cite{Madden20}, which were derived from galaxies with \LCII$ > 10^7$~L$_{\odot}$ and \LCII$ > 10^5$~L$_{\odot}$, respectively. 
The fit by \cite{Zanella18} applies to lensed galaxies at $z = 2$, although it is routinely used to infer $M_{ \rm H_2}$ from \LCII\ for higher-redshift galaxies as well.
This fit systematically lies above \coldsim\ haloes.
The fit by \cite{Madden20} was instead derived from local dwarf galaxies, which have lower metallicities and masses than more typical local galaxies and are usually considered a good analogue for higher-redshift objects \citep{Izotov21, Schaerer22, Telford23}. 
This fit agrees better with our predictions and with the conclusion that
the total $\rm H_2$ mass is well traced by \LCII.

Finally, we highlight that our physically motivated fits extend the empirical relations between \LCII\ and the mass of cold molecular gas to galaxies with a \CII\ luminosity \LCII $ < 10^7$ L$_{\odot}$, for which observations are still missing. Our fits also provide a reference for high-redshift observations.

\subsection{Conversion factors and galaxy properties}

\begin{figure}[]
\begin{center}
  \begin{tabular}{c}
    \includegraphics[width=0.484\textwidth]{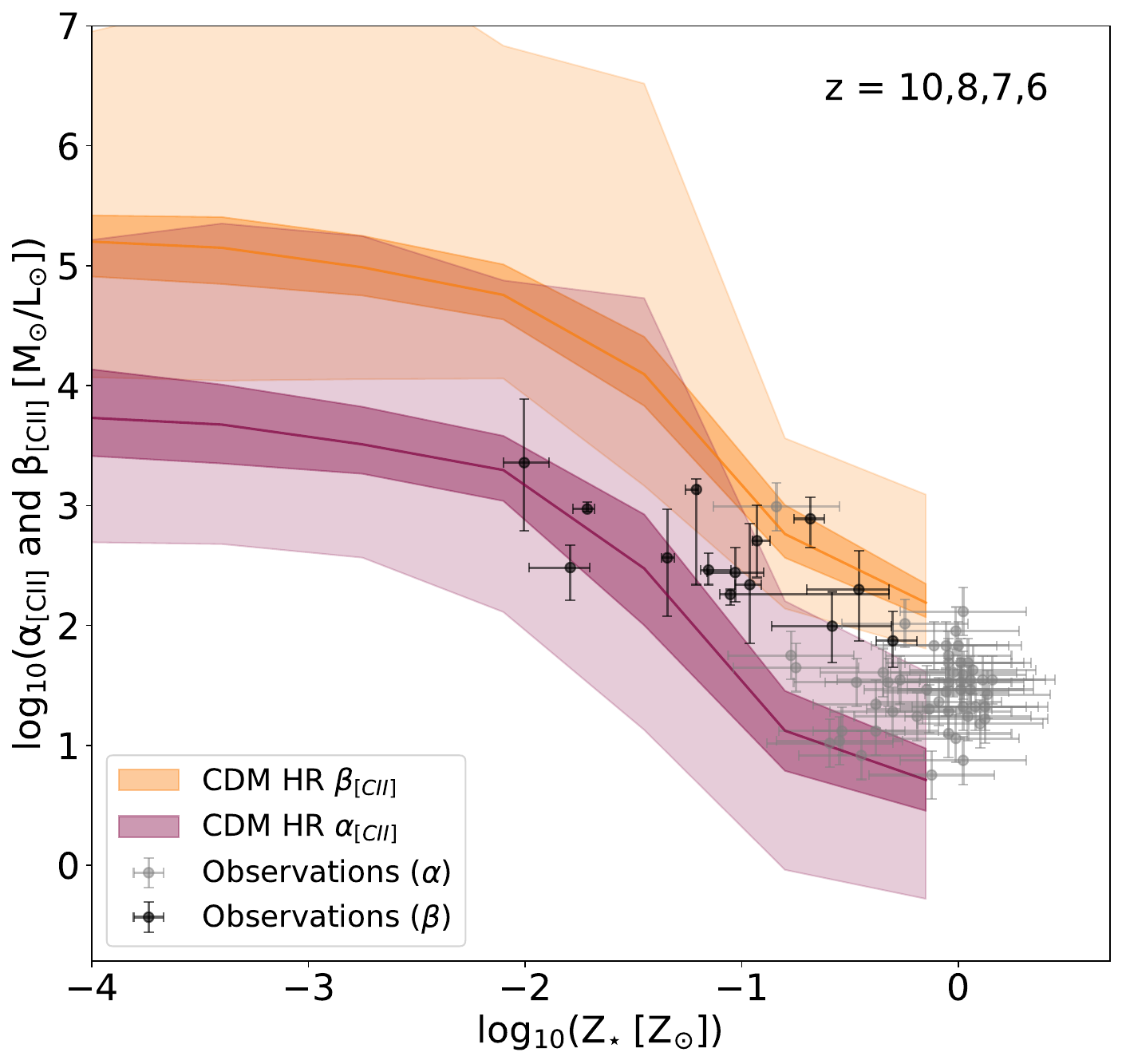}
  \end{tabular}
  \caption{\LCII-to-$M_{\rm H_2}$ conversion factor $\alpha_{\textup{[CII]}}$ (dark pink) and \LCII-to-$M_{\rm HI}$ conversion factor $\beta_{\textup{[CII]}}$ (orange) as a function of stellar metallicity $Z_{\star}$, at $z = 10, 8, 7$, and $6$ combined. The solid dark pink and orange lines represent the mean values obtained from CDM~HR for $\alpha_{\textup{[C II]}}$ and $\beta_{\textup{[C II]}}$, respectively. The darker (lighter) shaded regions represent the 1$\sigma$ (2$\sigma$) standard deviation. The grey points are $\alpha_{\textup{[C II]}}$ estimates from observations of local galaxies \citep{Cormier15, Accurso17, Contursi17, Hughes17} and $z = 2-4$ galaxies \citep{Huynh14, Schaerer15}. The black squares are $\beta_{\textup{[C II]}}$ derived by observations of  $\gamma$-ray burst afterglows in star-forming galaxies at  $z = 2-5$ \citep{Heintz21}. The \coldsim\ results agree well with lower-redshift observations and provide predictions for future observational works. Both $\alpha_{\textup{[C II]}}$ and $\beta_{\textup{[C II]}}$ increase by $\sim3 $ dex at low stellar metallicities.}
  \label{fig:alphavsZ}
\end{center}
\end{figure}

Based on the linear trend of the relations presented in section \ref{sect:linear relations}, we constrained the conversion factors \LCII-to-$M_{\rm H_2}$, $\alpha_{\textup{[C II]}}$, and \LCII-to-$M_{\rm HI}$, $\beta_{\textup{[C II]}}$. In this section, we explore their dependences on galactic properties such as $Z_{\star}$, $M_{\star}$, and SFR. In Figure \ref{fig:alphavsZ} we show the relation between $\alpha_{\textup{[C II]}}$ and $\beta_{\textup{[C II]}}$  and the galactic stellar metallicity $Z_{\star}$ at $z = 10,8,7$, and $6$ combined. $\alpha_{\textup{[C II]}}$ and $\beta_{\textup{[C II]}}$ are roughly constant for $Z_\star < 10^{-2}$ Z$_{\odot}$. In this regime, the scatter around the two relations is large because these low-mass haloes contain fewer gas particles.

\begin{figure*}[h]
\begin{center}
  \begin{tabular}{c c}
    \includegraphics[width=0.49\textwidth]{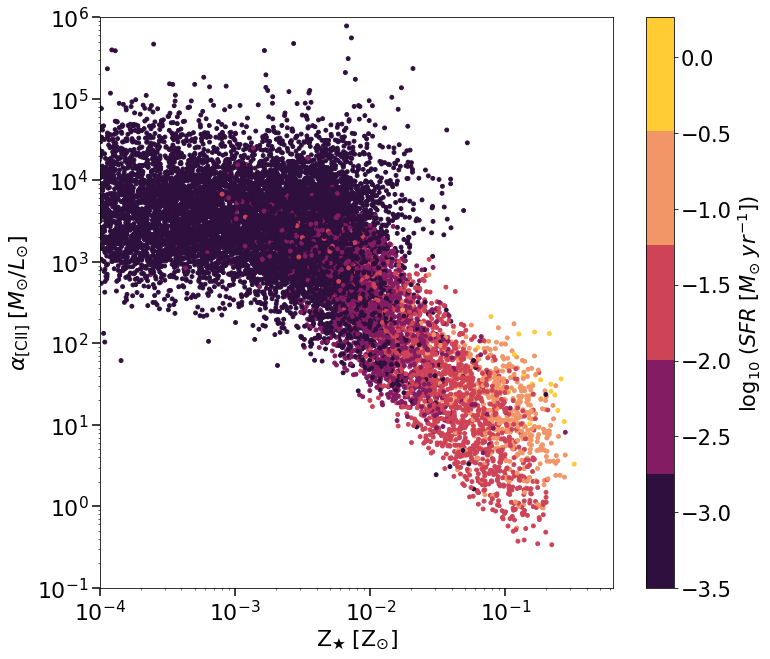} &
    \includegraphics[width=0.48\textwidth]{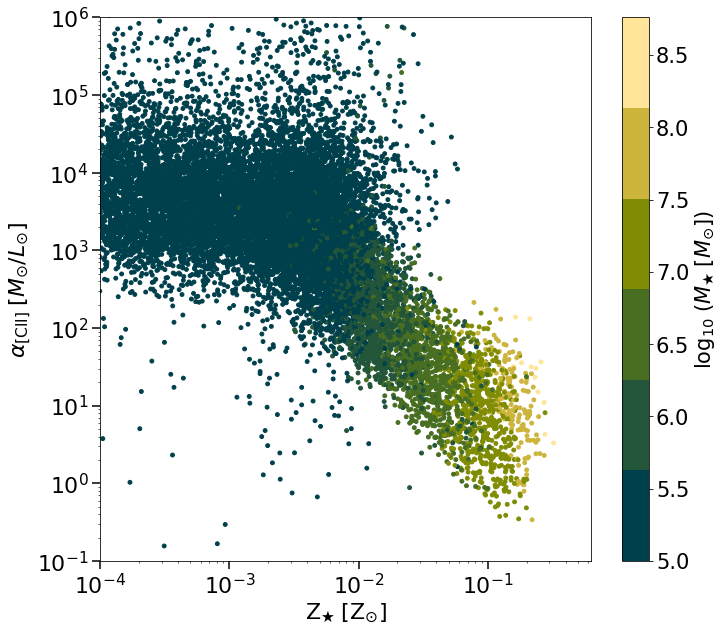}
  \end{tabular}
  \caption{Conversion factor $\alpha_{\textup{[C II]}}$ as a function of stellar metallicity $Z_{\star}$ colour-coded according to SFR (left panel) and stellar mass (right panel) for \coldsim\ galaxies. We highlight that the scatter around the relations presented in Figure \ref{fig:alphavsZ} is mainly affected by the SFR
  at $Z_{\star} > 10^{-2.2}$ Z$_{\odot}$.} 
  \label{fig:alphavsZ_sfrMstar}
\end{center}
\end{figure*}

In the range $Z_{\star} = (10^{-2}-10^{-1})$ Z$_{\odot}$, both factors experience a drop of about 2 dex on average. The reason for this is that these galaxies have experienced star formation activity (typically, $\rm SFR > 10^{-2}$ M$_{\odot}$  yr$^{-1}$), and stellar feedback has enriched the gas with metals. The haloes are therefore brighter in \LCII. When an increase in \CII\ luminosity is not followed by an equally efficient increase in neutral mass, $\alpha_{\textup{[C II]}}$ and $\beta_{\textup{[C II]}}$ 
may drop to even lower values. 
At $Z_{\star} > 10^{-1}$~Z$_{\odot}$, $\alpha_{\textup{[C II]}}$ and $\beta_{\textup{[C II]}}$ are again nearly constant, with a 2$\sigma$ scatter of$\sim 2$ dex for $\alpha_{\textup{[C II]}}$ and slightly larger than $1$ dex for $\beta_{\textup{[C II]}}$. This regime represents more massive galaxies that have experienced star formation episodes and significant metal enrichment. Since they are in a more evolved stage, they have had time to restore their neutral atomic and molecular gas content. this is reflected in the shallower $\alpha_{\textup{[C II]}}$ and $\beta_{\textup{[C II]}}$ trends.

For all the metallicities, a large 2$\sigma$ scatter of $\sim 5-3$ dex for $\alpha_{\textup{[C II]}}$ and $\beta_{\textup{[C II]}}$ indicates that it is difficult to constrain the mass of H$_2$ and \HI\ from \LCII precisely, as is also displayed in Figure \ref{fig:L_CIIvsMH2andMHI}.
Our estimates of $\alpha_{\textup{[C II]}}$ agree well with observations of galaxies with $Z_{\star} > 10^{-1}$ Z$_{\odot}$ in the range $z = 0-4$ \citep{Huynh14, Cormier15, Schaerer15, Accurso17, Contursi17, Hughes17}, especially with the local dwarf galaxies described in \cite{Cormier15}.
Moreover, we compared our results of $\beta_{\textup{[C II]}}$, with observations of star-forming galaxies at $z= 2-5$ by \cite{Heintz21}. The metallicities $Z_{\star} > 10^{-1.5}$ Z$_{\odot}$ agree welll.
At lower metallicities, the \coldsim\ simulations predict higher values of $\beta_{\textup{[C II]}}$. The reason might be that \cite{Heintz21} estimated $M_{\rm HI}$ from $\gamma$-ray bursts, which mainly probe the very dense gas ($N_{\rm H} > 10^{21}$ cm$^{-2}$), while our values also take the diffuse \HI\ retained in the galaxies into account. Early work from \cite{Vizgan22HI} reported a trend with metallicity that is consistent with the findings presented here. Our theoretical results can be useful for observational work in order to constrain the role that the metallicity plays in determining the trend of conversion factors. Especially for low-metallicity galaxies, our findings explore a regime in which observations struggle to characterise molecular and atomic gas masses.

To better understand the trend and scatter of $\alpha_{\textup{[C II]}}$ as a function of $Z_{\star}$, we show in Figure \ref{fig:alphavsZ_sfrMstar} the same relation, colour-coded according to SFR (left panel) and stellar mass (right panel). Galaxies with $Z_{\star} < 10^{-2}$ Z$_{\odot}$ typically have an SFR~$< 10^{-2.2}$~M$_{\odot}$~yr$^{-1}$ and $M_{\star} < 10^{6}$ M$_{\odot}$. These haloes are cold and poor in metals, and their \LCII\ is faint. 
For higher metallicities, however, the drop in $\alpha_{\textup{[CII]}}$ 
coincides with an increase in SFR and $M_{\star}$. As also demonstrated in C24, more massive and star-forming galaxies are also richer in metals and thus in carbon. This result agrees with the evolution of the mass-metallicity relation (C24), butwhile masses vary by one or two orders of magnitude, \LCII\ can vary by four or five dex (see e.g. previous Figure~\ref{fig:L_CIIvsMH2andMHI}).

The wide scatter of $\sim3$ dex at $Z_{\star}$ above $10^{-2}$ Z$_{\odot}$ for a fixed metallicity causes a larger scatter in SFR. 
For example, galaxies with a higher SFR exhibit higher values of $\alpha_{\textup{[C II]}}$ than quiescent galaxies in the same $Z_\star$ bin. 
This is a result of the stellar feedback, so that galaxies that form more stars are also hotter, with a mass-weighted temperature above $\sim 10^5$~K. Therefore, carbon in these haloes is present in higher ionisation degrees, and \LCII\ decreases. This effect implies that $\alpha_{\textup{[C II]}}$ values for galaxies with SFR $\sim 1$~M$_{\odot}$~yr$^{-1}$ are higher by up to 2~dex than those of galaxies with SFR $\sim 10^{-1.5}$~M$_{\odot}$~yr$^{-1}$ at a given $Z_{\star}$.
No such strong dependence of the scatter on the stellar masses is observed at fixed metallicity. This agrees with the tighter mass-metallicity relation discussed by C24 for example.
We additionally confirmed that the behaviour of $\alpha_{\textup{[C II]}}$ with the specific SFR shows no clear trend. This is consistent with the SFR and $ M_\star$ results shown in Figure~\ref{fig:alphavsZ_sfrMstar}.

\subsection{Gas phases traced by \CII\ emission}
\label{sect:gasphases}

\begin{figure}[h]
    \includegraphics[width=0.5\textwidth]{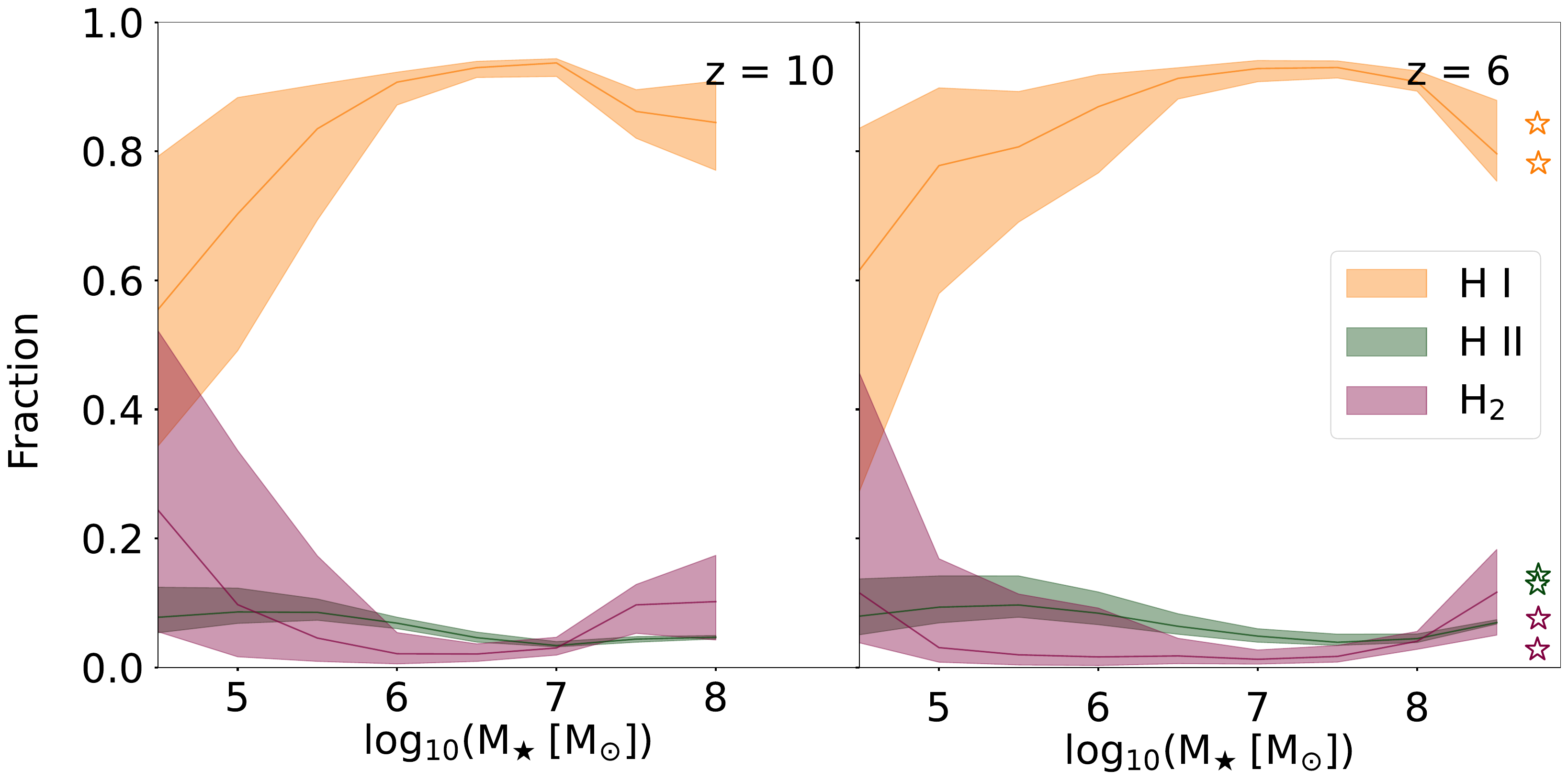} \\
    \includegraphics[width=0.5\textwidth]{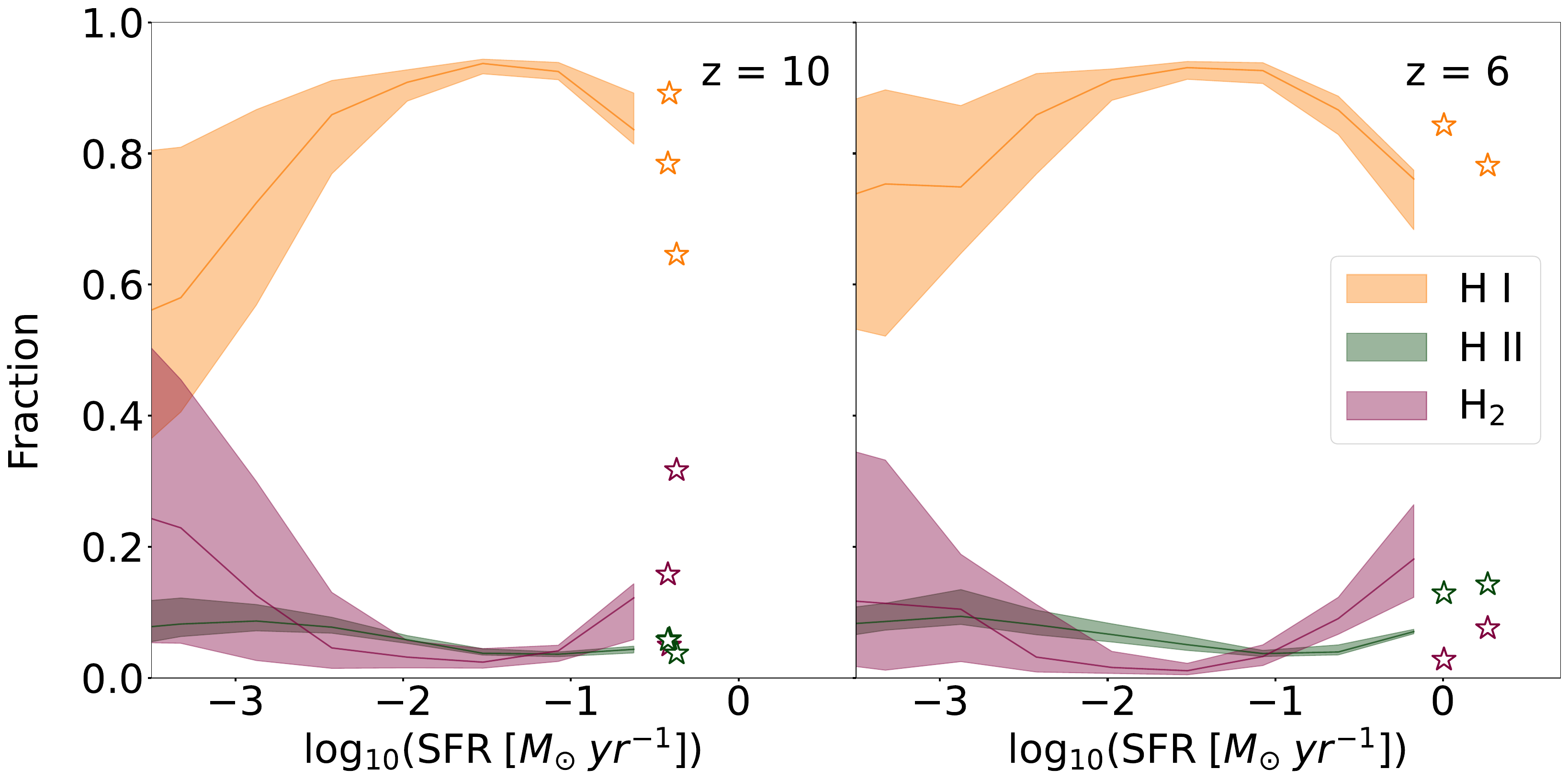}\\
    \includegraphics[width=0.5\textwidth]{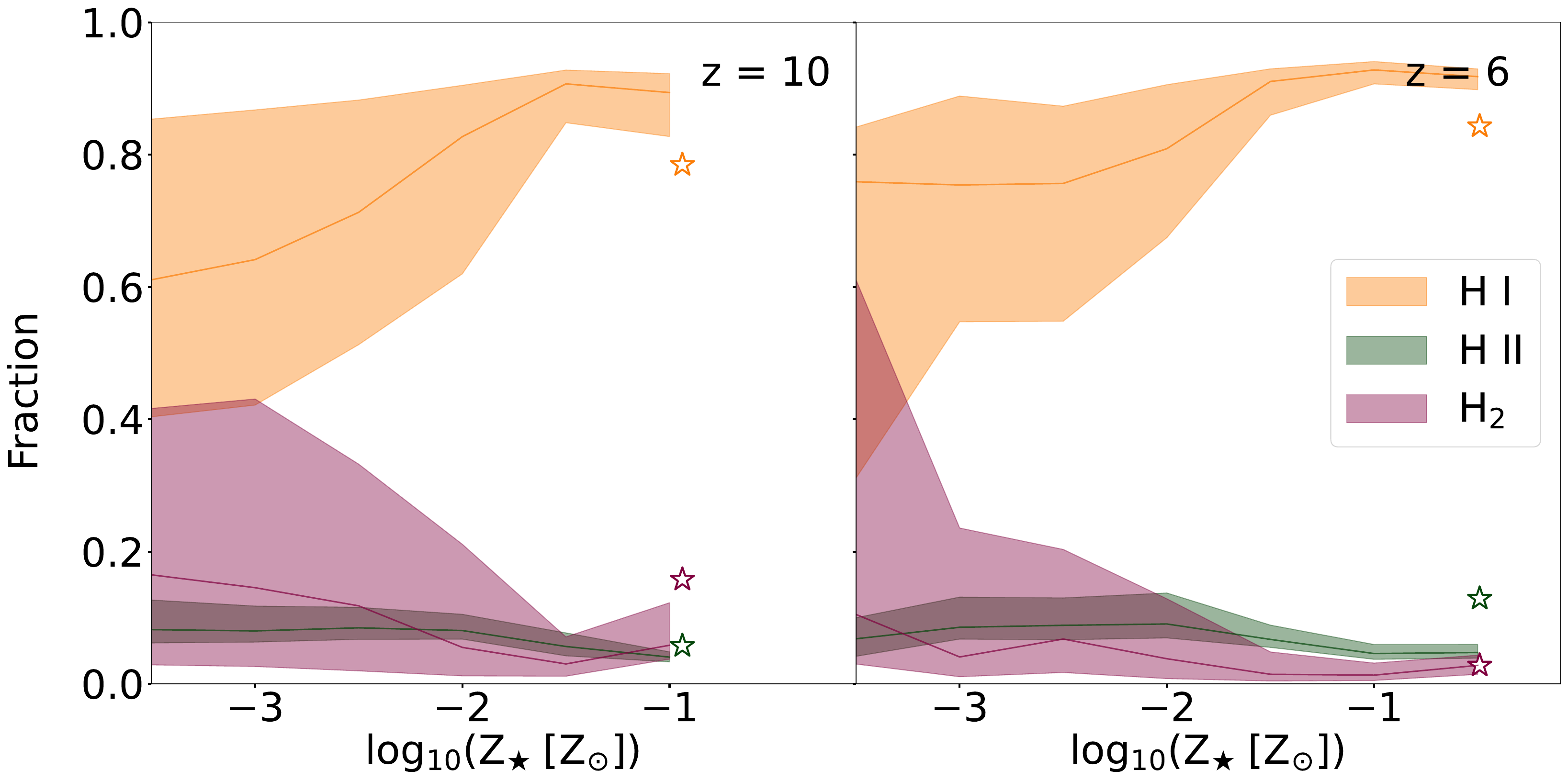} \\
    \includegraphics[width=0.5\textwidth]{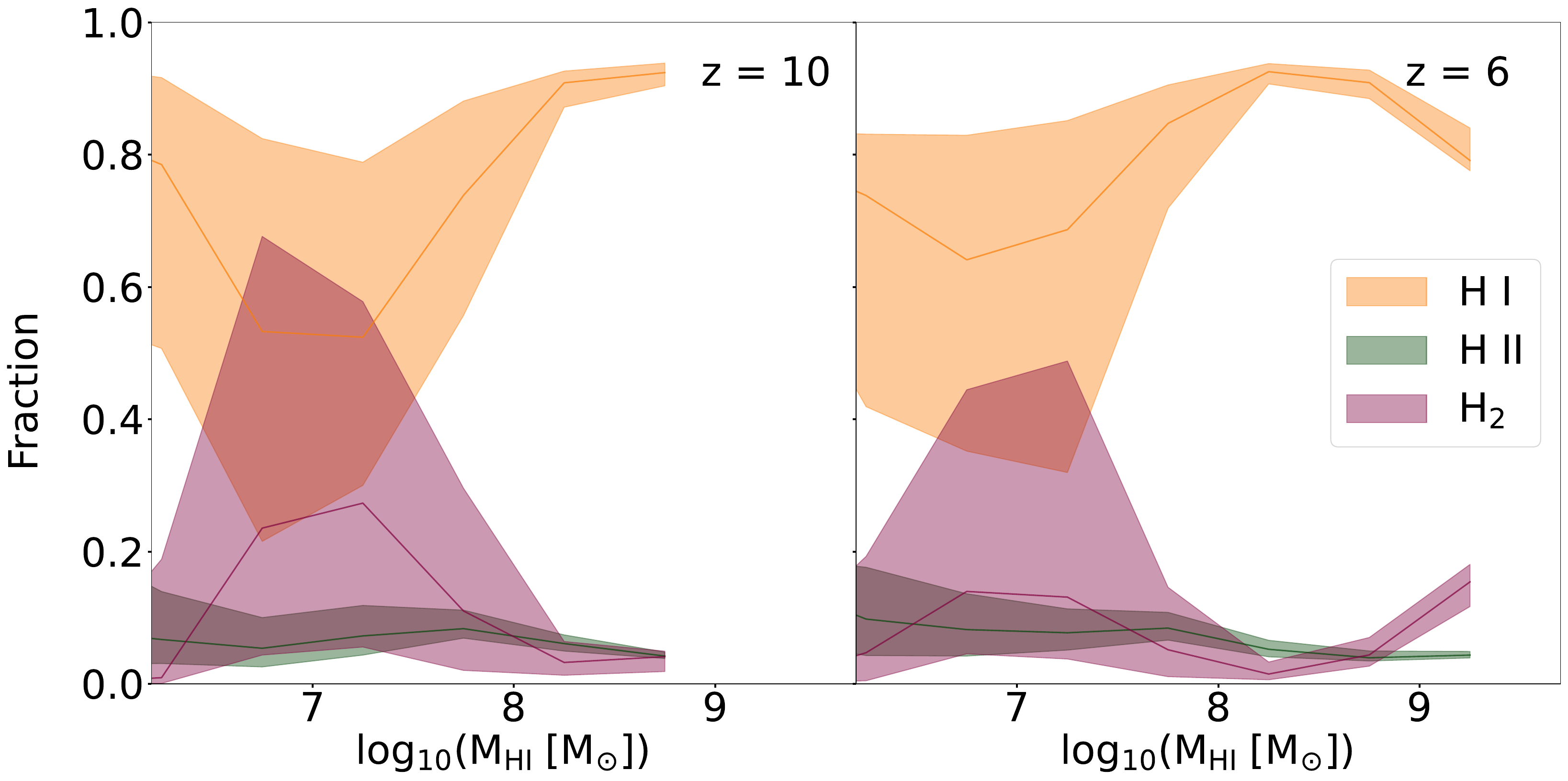}
  \caption{Fractional phase contribution to \CII\ luminosity as a function of galactic properties. From top to bottom: Contribution to \LCII\ from \HI\ (orange), H$_2$ (dark pink), and \HII\ (green) as a function of stellar mass, star formation rate, stellar metallicity, and \HI\ mass at $z=10$ (left panels) and 6 (right). The solid line displays the median values, and the shaded areas show the 1$\sigma$ scatter. The stars represent the contribution from the three phases in bins with fewer than three galaxies. The \CII\ emission mostly traces neutral atomic gas.}
  \label{fig:gasfractions}
\end{figure}

As mentioned in the previous sections, \LCII\ is a powerful tracer of cold gas in general.
In this regime, different gas phases may coexist, however. Cold gas typically hosts atomic (\HI) and molecular (H$_2$) hydrogen and residual ionised \HII\ gas.
As mentioned above, although at temperatures below few  $10^{4}\,\rm K$ the \HI\ mass is expected to be dominant with respect to the H$_2$ and \HII\ masses, it is unclear which gas phase is associated with most of the \CII\ emission because \LCII depends on the density and temperature.

The contribution to the \CII\ luminosity from each gas phase (ionised, molecular, and atomic) is shown in Figure \ref{fig:gasfractions} as a function of $M_{\star}$, SFR, $Z_{\star}$, and $M_{\rm HI}$.
In order to compute the contribution of each gas phase to the total \LCII\ of a galaxy, we decomposed the total \CII\ emission (eq.~\ref{eq:lum})  
into its components 
( see Appendix~\ref{sect:appendix})
for each gas particle, and we then summed the \LCII\ particle contributions in each galaxy. We specify that the contribution in luminosity arising from \HII\ gas is mainly due to interactions with electrons.

We note that at each redshift, most of the \CII\ emission is generated by collisions with \HI. However, the contribution of each phase varies with $M_{\star}$ and SFR.
At all redshifts, for stellar masses between $10^5$ and $10^{7.5}$~M$_{\odot}$ and SFR of $10^{-2}$ to $10^{-0.5}$~M$_{\odot}$~yr$^{-1}$, the contribution to \LCII\ from collisions with \HI\ is approximately $90\%$, and that of \HII\ can reach up to $10\%$, while the contribution from H$_2$ is often negligible. In practice, stellar feedback makes these galaxies brighter in \LCII until the gas particles reach temperatures higher than a few $10^4$~K.
Higher temperatures are reached in galaxies with $M_{\star} > 10^{7.5}$ M$_{\odot}$ and SFR $>10^{-0.5}$ M$_{\odot}$ yr$^{-1}$ (right panels in the figure). As a result, \LCII\ in these galaxies does not increase with stellar mass or SFR, and it therefore contributes less \HI\ than less massive and star-forming galaxies. This decrease in the \HI\ contribution is compensated for by an increase in the fraction of \LCII\ produced by H$_2$ (up to almost 30\%) instead of \HII\ (a few percent).
These haloes have high values of $M_{\rm H_2}$, as shown in Figure \ref{fig:L_CIIvsMH2andMHI}. By contrast, in gas particles with higher temperatures that are dominated by ionised gas, carbon is excited at higher ionisation levels. Haloes with $M_{\star} < 10^6$~M$_{\odot}$ and SFR $< 10^{-2}$~M$_{\odot}$~yr$^{-1}$ show a remarkable scatter. As already mentioned in previous sections, these small and quiescent galaxies are more susceptible to feedback effects, and this increases the scatter around the median values. However, it is challenging to estimate $M_{\star}$ and SFR in these galaxies observationally. Therefore, we show in the two lower panels of Figure \ref{fig:gasfractions} the trends for the contributions of the three phases to \LCII\ versus $Z_{\star}$ and $M_{\rm HI}$. The trends are the same as described above: \HI\ is the phase that dominates \LCII , with a fractional contribution $> 80 \%$ for $Z_{\star} > 10^{-2}$ Z$_{\odot}$ and $M_{\rm HI} > 10^8$ M$_{\odot}$. The trend with $M_{\rm HI}$ denotes large spreads at low HI masses, which corresponds to haloes with a lower gas particle number.

Our \coldsim\ simulations suggest that \LCII\ largely traces atomic \ion{H}{i} gas. This result is different from some interpretations of observations in the literature, which concluded that $\sim 60 - 80 \%$ of the \CII\ emission arises from H$_2$ gas \citep{Pineda13, DeBreuck19}. Our findings are a non-trivial consequence of the \CII\ emission properties and the fact that \HI\ is the dominant species in gas below $\sim 10^4 \,\rm K $, while H$_2$ is predominant in rarer cold and dense regimes. This result is valid at any redshift and stellar metallicity for galaxies with $M_{\star} > 10^5$ M$_{\odot}$ and SFR $> 10^{-2.5}$ M$_{\odot}$ yr$^{-1}$, and it agrees with observational estimates (e.g. \citealt{Croxall17, Cormier19}).
In particular, the recent results of \cite{Ikeda24} from the CRISTAL survey on 34 spatially resolved galaxies in the range $z = 4-6$ agree well with our results. The authors argued that the extended \CII\ emission in half of the sample originated in the diffuse medium, either \HI\ or \HII. We also emphasise that although \LCII\ largely traces the atomic \HI\ gas phase, this does not contradict our conclusions about the \LCII\ versus $M_{\rm H_2}$ relation in Section~\ref{sect:linear relations} and previous observational claims by \cite{Madden20} or \cite{Zanella18}, for instance.
\LCII\ can be a robust indicator of the mass of molecular gas, notwithstanding the typically smaller amounts of H$_2$ in this temperature regime. 
\CII\ correlates with cooling and clumping of gas, which in turn enhances H$_2$-driven star formation and increases collisional excitation, and consequently, emissivity.

\subsection{\CII\ relation to dense and star-forming gas}

In this section, we discuss the contribution of star-forming gas to the \CII\ emission and characterise how different regimes in terms of density and temperature contribute to the total \LCII\ in the simulated box.
This was done for the sake of simplicity by various works that did not explicitly decompose the gas into its molecular or atomic form, but into star-forming or non-star-forming and dense or diffuse gas.

In the first row of table \ref{table:SF and warm-dense} we display the fraction of that \LCII\ originated in star-forming parcels with an SFR~$ > 0 $ ~M$_{\odot}$~yr$^{-1}$ ($f_{\textup{[CII], SF}}$) at $z = 10$, 8, 7, and 6. Moreover, we divide gas particles into three different phases: diffuse (diff, with $n_{\rm H} <10$ cm$^{-3}$), cold dense (CD, with $n_{\rm H} >10$ cm$^{-3}$ and $T< 10^4$ K), and warm dense (WD, with $n_{\rm H} >10$ cm$^{-3}$ and $10^4$ K $ < T< 4 \times 10^4$ K). We chose $4 \times 10^4$ K as the upper limit for the WD phase since particles hotter than this do not emit in \CII\ (see C24 for more details). Based on these definitions, we then computed the fraction of \LCII\ from the different regimes (i.e. $f_{\textup{[CII], SF}}$,
$f_{\textup{[CII], diff}}$,
$f_{\textup{[CII], CD}}$, and $f_{\textup{[CII], WD}}$).
At each redshift, most of the \CII\ emission is generated by star-forming and WD gas, as displayed in table \ref{table:SF and warm-dense}.
This results is expected because star-forming particles typically have number densities above $10$ cm$^{-3}$ and are also close to WD particles because stellar feedback quickly increases the temperatures to above $10^4$ K. This result indicates that stellar feedback is essential for producing \CII\ emission in our simulations. On the one hand, gas is enriched with metals, including carbon, by SN explosions and AGB winds. On the other hand, this gas has the physical conditions in terms of density and temperature to excite \CII\ for the emission at 158$\,\mu$m. 
The increase in the contribution from WD and star-forming gas from $z = 10$ to $z = 6$ is in line with the increasing SFR density reported in C24. This shows that increasingly more gas in the simulation turns into stars. Moreover, in C24, we found that \CII\ can be used to trace star formation in the same redshift range.

This result might be sensitive to star formation and feedback prescriptions, metal enrichment details, and particle resolution. 
Different-resolution runs resolved gas regions that were dense enough to emit radiation via \CII\ 158~$\mu$m fine-structure line differently.
\cite{Schimek24}, for example, studied the fraction of \LCII\ that is emitted by dense particles with $n_{\rm H} > 10\,\rm cm^{-3}$ and $T < 10^4$ K and by diffuse gas particles at lower densities. They examined a zoom-in simulation of a single galaxy at redshift $z = 6.5$ and computed the \CII\ luminosity in post-processing, finding that most of the \LCII\ originated from dense particles, although a high percentage of $44\%$ is emitted by more diffuse gas. Recent results by \cite{Gurman24} showed that most of the \CII\ emission is produced by warm medium, which they defined as gas with $n_{\rm H} > 10$ cm$^{-3}$ and a temperature of $3 \times 10^3$~K $ < T < 3 \times 10^4$~K. They performed high-resolution hydrodynamical simulations of a patch of the interstellar medium (with a pixel resolution of about 2~pc) including H$_2$ chemistry and Type~II SN feedback, and post-processed their simulated gas cells to infer the carbon abundances and \LCII\ signal. They reported that most of the \CII\ luminosity originated in warm neutral medium and \HII\ regions.
This result is not affected by metallicity, but it might vary with the choice of the temperature upper limit and with a more detailed treatment of low-temperature metal cooling and feedback.

\begin{table}[t]
\caption{Fraction of \CII\ luminosity produced in gas particles 
with $\rm SFR > 0\, M_{\odot}\,yr^{-1}$ ($f_{\textup{[CII], SF}}$), 
in diffuse gas with $n_{\rm H}<10$ cm$^{-3}$ ($f_{\textup{[CII], diff}}$),
in cold dense gas with $n_{\rm H}>10$ cm$^{-3}$ and $T < 10^4$ K ($f_{\textup{[CII], CD}}$), and 
in warm dense gas with $n_{\rm H}>10$ cm$^{-3}$ and $10^4$ K $ < T < 4 \times 10^4$ K ($f_{\textup{[CII], WD}}$) 
at $z =  10, 8 ,7$, and $6$.
}
\label{tab:my-table}
\begin{tabularx}{\columnwidth}{X|Y|Y|Y|Y|}
                                & $z = 10$                        & $z = 8$                        & $z = 7$                        & $z = 6$                        \\ \hline
 $f_{\textup{[CII], SF}}$ &  $0.72$ &  $0.92$ & $0.96$ & $0.97$ \\ \hline
$f_{\textup{[CII], diff}}$                      & $0.36$                         & $0.04$                         & $0.04$                         & $0.04$                         \\ \hline
$f_{\textup{[CII], CD}}$                        & $0.12$                         & $0.04$                         & $0.04$                         & $0.02$                         \\ \hline
$f_{\textup{[CII], WD}}$ & $0.52$ & $0.92$ & $0.92$ & $0.94$ \\ \hline
\end{tabularx} %
\label{table:SF and warm-dense}
\end{table}

\section{Discussion}\label{sect:Discussion}

The conversion of molecular gas into stars is traced by the depletion time, which is based on the ratio of the H$_2$ gas mass and SFR. From a theoretical perspective, it is difficult to properly model the physical processes involving molecular gas. Various scales need to be considered, ranging from large-scale cosmological environments to interstellar physics on a parsec scale. In this context, our work is intended to understand the origin of the measured \LCII\ signals and to support observers, especially in interpreting the properties of galaxies detected with facilities such as ALMA and JWST.

In the following, we first discuss the assumptions and caveats of our \coldsim\ numerical simulations. We then compare our results with other numerical works in the literature, and we finally show their usefulness for observational studies in accurately estimating the molecular gas mass in high-$z$ galaxies.

\subsection{\coldsim\ assumptions and caveats}

Our simulations, like all others in the literature, relied on a number of assumptions that affect the quantitative results shown in this paper.
We stress that the adoption of a detailed non-equilibrium model for chemical abundance calculations, similar to the one used here, is crucial for capturing the processes that shape atomic and molecular gas, as highlighted by \cite{Dave20} or \cite{Hu21}, for instance.

Possible caveats caused by the choices in the set-up were extensively described by \cite{Maio22}, \cite{Maio23}, and C24, for example. There, we already discussed the main assumptions about the initial stellar metallicity, different models of explosive nucleosynthesis, stellar wind variations in the values of the carbon yields by up to 1 dex, and the atomic physics uncertainties in the collisional rates needed to calculate the  \CII\ luminosity.

Sources of uncertainty may arise from the way in which stellar feedback is implemented. 
In \coldsim\ , each stellar particle is considered as a stellar population with a Salpeter \citep{Salpeter55} initial mass function. Other IMF models, such as Chabrier \citep{Chabrier03}, which are more bottom heavy, would inject more $\alpha$ elements from supernova Type II explosions into the surrounding medium. This can be seen, for example, by considering the supernova Type II fraction of the two IMFs, which is $\sim 0.2 \%$ for a Salpeter and $\sim 3 \%$ for a Chabrier IMF.
In our study, the choice of the IMF might have an impact on \LCII because the mass of ejected carbon depends on the slope of the IMF. In contrast, we do not expect the choice of a Chabrier over a Salpeter IMF to have strong consequences on the values of $M_{ \rm HI}$ and $M_{ \rm H_2}$. \cite{Maio22} analysed both IMFs and did not note any significant differences in the primordial $\rm H_2$ and \HI\ cosmological mass density.

Our stellar feedback model assumes that
 Pop~III stars are distributed with a Salpeter IMF, similarly to
Pop~II and Pop~I stars.
However, different choices for the early Pop~III IMF are possible. For example, for a top-heavy Pop~III IMF, primordial stars would have much shorter lifetimes than Pop~II and Pop~I stars. Consequently, at high redshifts, they would more quickly enrich the interstellar medium with metals \cite[see results in e.g.][]{Maio10}, with possible implications for early \CII\ luminosities.

In addition to Type~II SNe, we also considered AGB stars and Type~Ia SNe, which are driven by lower-mass stars with longer lifetimes. Their typical timescales are about $10^8-10^9$~yr, and thus, they can enrich the cosmic medium already by $z\simeq6$.
This has direct implications for feedback effects and chemical abundances during the epoch of reionisation (as demonstrated by the first spectroscopic evidence of a mature stellar population in a $z\sim 8$ galaxy, A2744-YD4; \citealt{Witten24} and massive quenched galaxy at z $\sim 7$, RUBIES-UDS-QG-z7; \citealt{Weibel24}).
As an example, AGB stars produce relevant amounts of carbon in only few hundred million years, and therefore, the chemical gas patterns and the expected \CII\ emission may be severely affected by them.
The metal yields of Type~Ia SNe do not strongly depend on the initial stellar metallicity, but their occurrence requires timescales shorter than $10^8$
%
%
to more than $10^9$~yr and assumptions about the realisation probability of the Type~Ia SN scenario. This latter is basically unknown for early epochs. We therefore relied on standard prescriptions suggested by stellar evolution models \cite[for usual stellar parameters the delay time distribution peaks well below one Gyr;][]{Tinsley80, Greggio05, Maoz17}.
In this way, we were able to take their possible impact on our \CII\ luminosity estimates into account as well.
We stress that these concerns about AGB enrichment and Type~Ia SNe must be kept in mind when focusing on primordial metal signatures because most high-redshift numerical simulations tend to neglect them
(although roughly half Type~Ia SNe in a single stellar population explode within less than a billion years; \citealt{Cappellaro15}).
%
%

In our work, galactic winds take place at a fixed velocity of $350\,\rm km/s$ and are temporarily decoupled from cooling and star formation \cite[according to][]{Maio11}. This allowed us to obtain a more efficient mass ejection and to mitigate potential overcooling problems, and at the same time, to prevent unphysical behaviour related to the formation of stars in the wind phase.
The uncertainties about wind velocities (or mass-loading factors) are large, but we verified that at these early epochs, the exact adopted value has no drastic effects for the \HI\ and H$_2$ phases because the mass of the hosting haloes is relatively low \citep{Maio22}.
At intermediate and low redshifts, the impacts might be more relevant, however \cite{Nagamine04}, while a decrease in outflow velocity might lead to more prominent galactic blue peaks at $z > 6$ \citep{Hassan21}.

Throughout this paper, we did not consider feedback from primordial black holes (BHs). Their presence might cause a displacement in the gas that would create turbulence and heat the surrounding medium. 
In addition to the huge uncertainties in early-BH models, it is possible that this effect reduces the amount of cold gas, particularly in its molecular form, which would cause the baryons to be in a hotter phase. However, at these redshifts, the masses of BHs are expected to be lower than those in the local Universe, and if the BH mass-bulge mass relation still holds, their masses should be $\sim 10^{-3}$ times the hosting stellar masses at most (see detailed figures in C24). Severe implications from early-BH seeds would only be found in very extreme cases, such as those discussed in \cite{Woods24}, \cite{Patrick23}, and \cite{Maio19}. Thus, the inclusion of primordial BHs is not expected to have drastic consequences for our findings in general.

In \coldsim\, heavy elements produced by stars are spread over the neighbouring particles, and metal mixing is mimicked through SPH kernel smoothing. This means that in a low-density region at larger distances from the point of origin of the metals, the metallicity might be underestimated. This is typical of many SPH simulations, however, and winds are usually helpful to alleviate this problem.
Since we focused on cold-gas phases and particles close to star-forming regions, this effect is not expected to create significant consequences in our results.

Variations in the adopted cosmological parameters, higher-order corrections to linear theory, or in the assumed dark-matter scenario are likely to play a minor role because \CII\ emission essentially stems from cold gas that is bound within the structure potential wells and is not affected long-distance cosmological phenomena.
They might alter the high-redshift statistics, however \cite[for a deeper discussion on the topic see e.g.][and references therein]{Yoshida03, Maio06, Maio11L, Maio15, Maio23}.

\subsection{Comparison with other numerical studies}

Numerically, it is one of the greatest challenges to accurately model physical processes that cause the \CII\ fine-structure line emission at 158 $\mu$m in a cosmological context. \coldsim\ has a non-equilibrium chemical network implemented on the fly that is designed to describe the primordial gas chemistry. Specifically, at each time step, the abundances of each atomic and molecular gas species are computed in a consistent way with cooling and heating processes. 
Moreover, individual elements, including carbon, are tracked separately and are not inferred from a global metallicity. This means that the amount of carbon in each particle is estimated at each simulation time step according to the underlying stellar evolution model. 
This dedicated set-up enables the calculation of the contribution of each gas phase, ionised, atomic, and molecular, to the \CII\ emission.

As the results discussed in this paper are sensitive to the implementation adopted to track the different gas phases, we compared our findings with those from other numerical work in the literature. We emphasise that the various models differ in terms of initial set-up, box size, resolution, feedback implementation, sub-grid physics, and post-processing of \LCII\ and gas-phase contributions.

Previous works investigated the contribution of different gas phases to the \CII\ emission in galaxies at high $z$ via semi-analytical models \citep{Vallini15, Lagache18, Popping19, Yang22}, zoom-in simulations of individual galaxies (\citealt{Olsen17, Olsen18, Pallottini19, Lupi20, Bisbas22, Schimek24, Khatri24a}), or a section of the interstellar medium \citep{Gurman24}, and post-processing of cosmological simulations (\citealt{RamosPadilla21, RamosPadilla23, Vizgan22H2, Vizgan22HI, Garcia23}).
All these investigations were based on a number of different assumptions and sub-grid models. This can lead to large discrepancies in the results.
For example, with the exception of \cite{Olsen17, Olsen18}, \cite{Lupi20}, and \cite{Bisbas22}, all other works traced a global metallicity and inferred the carbon mass in post-processing. \coldsim\ instead self-consistently follows individual elements according to stellar evolution and the associated gas cooling. 
Additionally, as it is computationally expensive to include H$_2$ processes in the chemical network as here, only a few works included them, and for single haloes, they either implemented \CII\ emission as well (\citealt{Bisbas22}) or neglected it (\citealt{Pallottini17, Pallottini19}).

Only a handful of studies investigated the contribution of different gas phases to \LCII\ at the epoch of reionisation through cosmological simulations. All these works estimated the \HI\ and H$_2$ masses in post-processing, however. Here, we mention \cite{Vizgan22H2, Vizgan22HI, Garcia23} with SIMBA, and \cite{RamosPadilla21,  RamosPadilla23} with EAGLE. They covered box sizes ranging from $25$ cMpc/$h$ to $100$ cMpc/$h$.

As described in Section \ref{sect:linear relations}, our results robustly indicate that the logarithms of \LCII\ and those of $M_{\rm H_2}$ and $M_{\rm HI}$ are linearly correlated. 
Previously, this relation was only studied via \CII\ post-processing modelling by \cite{Vizgan22H2, Vizgan22HI}. The slope and amplitude of their relations are different from ours, but the range of masses and luminosities from which our fits  and the SIMBA fits were derived are different. 

Another important result is the evolution with redshift of the different galaxy properties, including \LCII, molecular and atomic gas mass, SFR, metallicity, and stellar mass. 
In this respect, \cite{Liang24} studied the empirical \LCII-SFR relation with the FIRE cosmological zoom-in simulations. They focused on the so-called \CII\-deficit galaxies \citep{Knudsen16}.
The authors reported two distinct physical regimes: H$_2$-rich galaxies, in which the depletion time is the main driver of the \CII\ deficit; and H$_2$-poor galaxies, in which gas metallicity dominates.
Consistently with C24, their findings suggested that the \CII\ deficit is a widespread characteristic of galaxies, highlighting the need for caution for using a constant \LCII-to-SFR conversion factor (derived from local star-forming galaxies) to estimate SFRs at high redshift \citep[see, e.g.,][]{Righi08, Sommovigo2021, Roy24, Khatri24b}.

In our study, we investigated the dependence of the conversion factors  $\alpha_{\textup{[CII]}}$ and  $\beta_{\textup{[CII]}}$ on galactic properties. We found that the two conversion factors are dependent on the stellar metallicity, $Z_{\star}$. Additionally, for $Z_{\star}  > 10^{-2}$~Z$_{\odot}$, galaxies with larger $\alpha_{\textup{[CII]}}$ are those with a higher SFR, which shows that the conversion factor depends strongly on the star formation activity in the galaxy.

In Section~\ref{sect:gasphases} we showed that \HI\  gas phase dominates \CII\ emission, while the H$_2$ gas phase can contribute no more than $\sim 30\%$ to a galaxy \LCII.
Thus, some of the previous findings \citep[such as the ones by][]{Vallini15, Lupi20, RamosPadilla21, RamosPadilla23, Bisbas22, Gurman24} agree with ours.

For the sake of simplicity, a number of works did not explicitly decompose the gas into molecular or atomic gas, but into star-forming or non-star-forming and dense or diffuse gas \citep{Schimek24, Gurman24}. \cite{DiCesare24} recently speculated that in galaxy mergers at $z > 4$, about $50\%$ of the \CII\ emission might come from the diffuse medium that originated in the middle of the merger. The authors proposed that even at lower redshifts, \CII\ emission from diffuse gas is not negligible. 
 \cite{Munoz24} studied the \CII\ emission in a sample of 357 galaxies at redshift $z=5$. Despite the limited simulation resolution and possible shortcomings in the adopted modelling, they suggested that the majority of \LCII\ originated from dense star-forming sites ( assuming a star formation density of $0.1\,\rm cm^{-3}$ and a fixed cold-phase gas temperature of 1~000~K), while the \CII\ extended emission mainly stems from satellite galaxies 
within the virial radius.
We note again that different simulations and post-processing methods provide non-converging results.

In addition to these numerical efforts that attempted to characterise $M_{\rm H_2}$ with \LCII, others used the CO emission. Among them, \cite{Hu23} presented high-resolution ($\sim$0.2~pc) hydrodynamical simulations of an isolated low-metallicity (0.1~Z$_{\odot}$) dwarf galaxy coupled with a time-dependent chemistry network. This metallicity regime is of particular relevance to the high-redshift objects that are now unrevealed by JWST \cite[][]{CurtisLake23}.
On a cosmological scale, \cite{Keating20} used the FIRE zoom-in simulations to reproduce the $L_{{\rm CO}}$-to-$M_{\rm H_2}$ conversion factor. 
They relied on an equilibrium solver for estimating H$_2$ and CO abundances in the post-processing and reported little evolution of the $L_{{\rm CO}}$-to-$M_{\rm H_2}$ conversion factor with SFR. This contradicts our non-equilibrium findings about the \LCII-to-$M_{\rm H_2}$ conversion factor.

\subsection{Predictions for high-$z$ observations}

Observationally, many works reported a linear relation between \LCII\ and the mass of H$_2$ in the local Universe and then estimated a conversion factor. Nevertheless, it is complicated to estimate it empirically at high redshift or at low metallicities. In these regimes, it becomes challenging to observe metal emission lines or to infer the mass of molecular gas without relying on \LCII. 
The maximum redshift reached to estimate $\alpha_{\textup{[CII]}}$ from galactic gas is $z \simeq 4$ \citep{Huynh14}, but the hosting objects are usually very luminous (\LCII $> 10^9$ L$_{\odot}$), massive in H$_2$, and are not statistically representative of the majority of high-redshift galaxies, which are less luminous and have lower masses. 
For galaxies at the epoch of reionisation, when CO cannot be observed, the conversion factor of
$\alpha_{\textup{[CII]}} = 31~ \rm M_{\odot}/L_{\odot}$
empirically derived by \cite{Zanella18} from star-forming lensed galaxies at $z = 2$ is often applied.  \cite{Dessauges-Zavadsky20, Aravena24}, and \cite{Kaasinen24} used this value to estimate the H$_2$ mass of galaxies in the redshift range $z = 4 - 8$.
 \cite{Kaasinen24} reported that estimates of $M_{\rm H_2}$ can vary by more than one dex depending on whether \CII\ or CO emissions are used as tracers. Recently, \cite{Salvestrini24}, demonstrated that conversion factor calibrated in the local Universe overestimate significantly the $M_{\rm H_2}$ measurement at the epoch of reionisation.
Moreover, \cite{Glazer24} reported that low-mass galaxies at $z \simeq 7$ present a deficit in \CII\ luminosity when their \LCII\ is compared to the expectations by the \LCII-SFR relation calibrated on local dwarf galaxies \cite[e.g.][]{DeLooze14}.
The authors justified this conclusion by invoking the low metallicities of their sample compared to those of the local Universe. This results therefore indicates that galaxies at high $z$ might have different properties from those at lower redshift and would require a physically motivated $\alpha_{\textup{[CII]}}$  calibration.

Our findings highlight that \LCII\ is dominated by the \HI\ atomic phase of the gas in high-redshift galaxies. Nonetheless, it is important to note that this result does not imply that all available \HI\ interacts with carbon and produces \LCII.
The \HI\ gas is clearly expected to be more extended than the \CII\ region, which requires ionising photons with energy greater than 11.3~eV, as also observed in nearby galaxies \citep{Peroux19, Szakacs21}. 
Moreover, pristine gas, such as the gas that is far from star-forming regions, could be rich in \HI, but contain no metals (hence would not emit any \LCII\ signal).
This is particularly true in primordial galaxies, when the amount of pristine material is expected to be greater than in the local Universe. We therefore need a \LCII-to-$M_{\rm HI}$ conversion factor $\beta_{\textup{[CII]}}$ computed at $z > 6$ that evolves with metallicity.

Our work is useful for estimating the mass of cold gas, atomic and molecular, in galaxies at the epoch of reionisation, such as those of the ALPINE and REBELS surveys. Furthermore, it would be interesting to understand how \CII\ emission is spatially distributed with respect to \HI\ and H$_2$. This type of analysis will be very useful in view of the new results from the CRISTAL survey, for example. 
The CRISTAL project targets 34 galaxies in the range $z = 4-6$ and spatially resolves their \CII\ emission.
Our simulations so far agree with recent results by \cite{Ikeda24} and \cite{Juno24},  who reported extended \CII\ emission that was primarily linked to star formation activity, even at kiloparsec scales, with contributions from atomic gas and mergers that might further expand the \CII\ line distribution.

Finally, in order to determine the observability of the simulated objects, we computed the exposure time needed for detecting \coldsim\ galaxies with the ALMA 12~m array. For this purpose, we used the CASA sensitivity calculator \citep{McMullin07} and determined the time required to detect the most luminous galaxies in our sample at 5$\sigma$.
We selected \CII\ emission at $z = 6$, corresponding to an observed frequency of $271.5$ GHz, which is observable in ALMA Band~7. To detect a galaxy with \LCII\ $\sim 10^7 \,\rm L_{\odot}$, the exposure time needed is seven hours on-source, or shorter for lower redshifts. Our predictions indicate that although it is still challenging to probe these galaxies, with the physical properties predicted by \coldsim, more massive haloes are within reach for ALMA.

\section{Conclusions}
\label{sect:Conclusions}

It is not yet possible to observationally resolve the spatial distribution of cold gas and to estimate the masses of \HI\ and H$_2$ components in galaxies at the epoch of reionisation. An important tool for tracing atomic and molecular gas at these redshifts is the fine-structure 158~$\mu$m \CII\ emission line observed by dedicated  facilities such as ALMA, and it is within the reach of the upcoming AtLAST project.
However, it is still unclear which phase of the gas at these epochs mostly causes the \CII\ luminosity, whether ionised, neutral atomic, or molecular, and what the high-$z$ trends of the corresponding mass-to-light ratios are.

In this work, we have extended our previous numerical analysis \cite[][]{Casavecchia24} and explored the processes that cause \CII\ emission by using selected runs from the \coldsim\ numerical simulations \citep{Maio22, Maio23, Casavecchia24}.
These cosmological simulations include in addition to the usual implementation of gravity and hydrodynamic calculations a physics-rich and accurate non-equilibrium chemical network that follows the evolution of atomic and molecular species, in particular, those we focused on here: \HI, \HII\, and H$_2$. Primordial chemistry was computed consistently with star formation and feedback effects and with cooling and heating processes.
With \coldsim\, , we tracked the time evolution of individual heavy elements, including carbon, and consequently derived the \CII\ emission from collision processes with \ion{H}{i}, \ion{H}{ii}, and H$_2$. 
In this way, we were able to provide physically motivated conversion factors that link the \CII\ luminosity, \LCII, to the H$_2$ mass ($\alpha_{\textup{[C II]}}$) and \HI\ mass ($\beta_{\textup{[C II]}}$), and their scaling with stellar mass and SFR.

Our main findings are summarised below.

\begin{enumerate}
    \item The \coldsim\ results reproduce the \CII\ luminosity function trends observed by ALMA. Our galaxies extend the \CII\ luminosity function to \LCII\ lower than in accessible to observations at $z = 6$ and provide first predictions in a range in which data are still limited.
    \item We found a linear relation between $\log_{10}($\LCII) and $\log_{10}(M_{\rm H_2})$ at all redshifts. Our results at $z = 6$ agree with observational estimates from galaxies in the range $z=0-7$. In particular, 
    the model by \cite{Madden20}, derived from local dwarf galaxies, broadly agrees with our estimates. The range of $\rm M_{H_2}$ and \LCII\ that is analysed and/or observed is not always the same, and thus, the comparison must refer to an extrapolation of the results.
    \item We also studied the relation between $\log_{10}($\LCII) and $\log_{10}(M_{\rm HI})$. We found that for a fixed \LCII\, the mass of \HI\ is typically $\sim$2~dex higher than that of H$_2$.
    \item The amplitude and scatter of the mentioned relations evolve with redshift. We provided redshift-dependent fitting formulae that link \LCII\ to $M_{\rm H_2}$ and $M_{\rm HI}$, and that can be used to estimate the mass of molecular and atomic gas of a galaxy from its (detected) \CII\ luminosity.
    \item We investigated the dependence of the $\alpha_{\textup{[C II]}}$ and $\beta_{\textup{[C II]}}$ conversion factors on the stellar metallicity, $Z_\star$, finding a good agreement with observational estimates at lower redshift.
    For $Z_\star<10^{-1}$ Z$_{\odot}$, the lack of carbon causes an increase of about 2~dex in the values of $\alpha_{\textup{[C II]}}$ and $\beta_{\textup{[C II]}}$, and this might help explain the common \CII\ deficit in galaxies around $z\simeq 7$ \cite[consistently with  observational findings by][]{Glazer24}.
    \item We explored the scatter in the relation between $\alpha_{\textup{[CII]}}$
    and $Z_{\star}$, finding that for metallicities $Z_{\star} > 10^{-2}$ Z$_{\odot}$, the large scatter depends on the galaxy SFR. Highly star-forming galaxies ionise carbon at higher ionisation states, making them less luminous in \LCII\ than galaxies with similar metallicity, but lower SFR.
    \item We examined the contribution of each gas phase, ionised, neutral atomic, and molecular, to the total emission in \LCII, and we studied the correlation with SFR and $M_{\star}$. We found that for an SFR~$> 10^{-2.5}$ M$_{\odot}$ yr$^{-1}$ and $M_{\star} > 10^5$ M$_{\odot}$, the phase that dominates \LCII\ is the atomic phase, with fractional contributions ranging from $70$ to $90\%$.
    \item The contribution from molecular gas varies, with values up to $\sim$30\% for galaxies with SFR~$\gtrsim 10^{-1}$~M$_{\odot}$~yr$^{-1}$ and $M_{\star} \gtrsim 10^8$~M$_{\star}$. Even though \LCII\ mainly traces \HI\ gas, it is still extremely useful to estimate the mass of the sub-dominant molecular phase because the fluctuations of H$_2$ the contributions across the considered mass range are relatively minor.
    \item We investigated the relation between the three gas-phase contributions to the total \LCII\ with $Z_{\star}$ and $M_{\rm HI}$. In this case, \HI\ is also the dominating phase, with a fractional contribution $> 80 \%$ for $Z_{\star} > 10^{-2}$ Z$_{\odot}$ and $M_{\rm HI} > 10^8$ M$_{\odot}$.
    \item In the entire simulated volume, the vast majority of \LCII\ is predicted to originate in warm dense and star-forming regions, regardless of the redshift. This agrees with the latest observational results by \cite{Ikeda24}.
\end{enumerate}

\noindent
The work presented here offers new theoretical insights to study the processes at the origin of \CII\ $158 \,\rm \mu m$ fine-structure line emission during the epoch of reionisation and supports recent and upcoming observations of star-forming gas in primordial times.
However, to better link theory and observations future investigations are still required. They will be crucial to reach a full understanding of carbon chemical signatures in the first Gyr and interpret correctly the origin of the first galaxies at such primeval epochs.

\begin{acknowledgements}
We acknowledge the anonymous referee for the constructive report, that allowed us to greatly improve the original manuscript.
We are thankful to Aniket Bhagwat for fruitful discussions on emission processes in high-$z$ galaxies. We are grateful to Rodrigo Herrera-Camus for providing a preview of the results on \CII\ luminosity in the CRISTAL galaxies. BC thanks Victoria Bollo for the help in the CASA calculations and the interesting discussion about molecular gas emission and absorption features through cosmic time. 
UM acknowledges financial support from the theory grant no.~1.05.23.06.13 ``FIRST -- First Galaxies in the Cosmic Dawn and the Epoch of Reionization with High Resolution Numerical Simulations'' and the travel grant no.~1.05.23.04.01 awarded by the Italian National Institute of Astrophysics.
UM was supported in part by grant NSF PHY-2309135 to the Kavli Institute for Theoretical Physics (KITP), Santa Barbara (USA).
This research was supported by the International Space Science Institute (ISSI) in Bern (Switzerland), through ISSI International Team project no.~564 (The Cosmic Baryon Cycle from Space). 
The numerical calculations done throughout this work have been performed on the machines of the Max Planck Computing and Data Facility of the Max Planck Society, Germany.
We acknowledge the NASA Astrophysics Data System for making available their bibliographic research tool
\end{acknowledgements}

\bibliographystyle{aa}
\bibliography{main}

\begin{thebibliography}{161}
\expandafter\ifx\csname natexlab\endcsname\relax\def\natexlab#1{#1}\fi

\bibitem[{{Abel} {et~al.}(1997){Abel}, {Anninos}, {Zhang}, \&
  {Norman}}]{Abel97}
{Abel}, T., {Anninos}, P., {Zhang}, Y., \& {Norman}, M.~L. 1997, \na, 2, 181

\bibitem[{{Accurso} {et~al.}(2017){Accurso}, {Saintonge}, {Catinella},
  {Cortese}, {Dav{\'e}}, {Dunsheath}, {Genzel}, {Gracia-Carpio}, {Heckman},
  {Jimmy}, {Kramer}, {Li}, {Lutz}, {Schiminovich}, {Schuster}, {Sternberg},
  {Sturm}, {Tacconi}, {Tran}, \& {Wang}}]{Accurso17}
{Accurso}, G., {Saintonge}, A., {Catinella}, B., {et~al.} 2017, \mnras, 470,
  4750

\bibitem[{{Andreani} {et~al.}(2020){Andreani}, {Miyamoto}, {Kaneko}, {Boselli},
  {Tatematsu}, {Sorai}, \& {Vio}}]{Andreani20}
{Andreani}, P., {Miyamoto}, Y., {Kaneko}, H., {et~al.} 2020, \aap, 643, L11

\bibitem[{{Aravena} {et~al.}(2016){Aravena}, {Decarli}, {Walter}, {Bouwens},
  {Oesch}, {Carilli}, {Bauer}, {Da Cunha}, {Daddi}, {G{\'o}nzalez-L{\'o}pez},
  {Ivison}, {Riechers}, {Smail}, {Swinbank}, {Weiss}, {Anguita}, {Bacon},
  {Bell}, {Bertoldi}, {Cortes}, {Cox}, {Hodge}, {Ibar}, {Inami}, {Infante},
  {Karim}, {Magnelli}, {Ota}, {Popping}, {van der Werf}, {Wagg}, \&
  {Fudamoto}}]{Aravena16}
{Aravena}, M., {Decarli}, R., {Walter}, F., {et~al.} 2016, \apj, 833, 71

\bibitem[{{Aravena} {et~al.}(2024){Aravena}, {Heintz}, {Dessauges-Zavadsky},
  {Oesch}, {Algera}, {Bouwens}, {da Cunha}, {Dayal}, {De Looze}, {Ferrara},
  {Fudamoto}, {Gonzalez}, {Graziani}, {Hygate}, {Inami}, {Pallottini},
  {Schneider}, {Schouws}, {Sommovigo}, {Topping}, {van der Werf}, \&
  {Palla}}]{Aravena24}
{Aravena}, M., {Heintz}, K., {Dessauges-Zavadsky}, M., {et~al.} 2024, \aap,
  682, A24

\bibitem[{{Bakes} \& {Tielens}(1994)}]{Bakes94}
{Bakes}, E.~L.~O. \& {Tielens}, A.~G.~G.~M. 1994, \apj, 427, 822

\bibitem[{{B{\'e}thermin} {et~al.}(2020){B{\'e}thermin}, {Fudamoto}, {Ginolfi},
  {Loiacono}, {Khusanova}, {Capak}, {Cassata}, {Faisst}, {Le F{\`e}vre},
  {Schaerer}, {Silverman}, {Yan}, {Amorin}, {Bardelli}, {Boquien}, {Cimatti},
  {Davidzon}, {Dessauges-Zavadsky}, {Fujimoto}, {Gruppioni}, {Hathi}, {Ibar},
  {Jones}, {Koekemoer}, {Lagache}, {Lemaux}, {Moreau}, {Oesch}, {Pozzi},
  {Riechers}, {Talia}, {Toft}, {Vallini}, {Vergani}, {Zamorani}, \&
  {Zucca}}]{Bethermin20}
{B{\'e}thermin}, M., {Fudamoto}, Y., {Ginolfi}, M., {et~al.} 2020, \aap, 643,
  A2

\bibitem[{{Bisbas} {et~al.}(2022){Bisbas}, {Walch}, {Naab}, {Lah{\'e}n},
  {Herrera-Camus}, {Steinwandel}, {Fotopoulou}, {Hu}, \&
  {Johansson}}]{Bisbas22}
{Bisbas}, T.~G., {Walch}, S., {Naab}, T., {et~al.} 2022, \apj, 934, 115

\bibitem[{{Blitz} \& {Rosolowsky}(2006)}]{Blitz06}
{Blitz}, L. \& {Rosolowsky}, E. 2006, \apj, 650, 933

\bibitem[{{Bolatto} {et~al.}(2013){Bolatto}, {Wolfire}, \& {Leroy}}]{Bolatto13}
{Bolatto}, A.~D., {Wolfire}, M., \& {Leroy}, A.~K. 2013, \araa, 51, 207

\bibitem[{{Boogaard} {et~al.}(2023){Boogaard}, {Decarli}, {Walter}, {Wei{\ss}},
  {Popping}, {Neri}, {Aravena}, {Riechers}, {Ellis}, {Carilli}, {Cox}, \&
  {Pety}}]{Boogaard23}
{Boogaard}, L.~A., {Decarli}, R., {Walter}, F., {et~al.} 2023, \apj, 945, 111

\bibitem[{{Bouwens} {et~al.}(2022){Bouwens}, {Smit}, {Schouws}, {Stefanon},
  {Bowler}, {Endsley}, {Gonzalez}, {Inami}, {Stark}, {Oesch}, {Hodge},
  {Aravena}, {da Cunha}, {Dayal}, {de Looze}, {Ferrara}, {Fudamoto},
  {Graziani}, {Li}, {Nanayakkara}, {Pallottini}, {Schneider}, {Sommovigo},
  {Topping}, {van der Werf}, {Algera}, {Barrufet}, {Hygate}, {Labb{\'e}},
  {Riechers}, \& {Witstok}}]{Bouwens22}
{Bouwens}, R.~J., {Smit}, R., {Schouws}, S., {et~al.} 2022, \apj, 931, 160

\bibitem[{{Capak} {et~al.}(2015){Capak}, {Carilli}, {Jones}, {Casey},
  {Riechers}, {Sheth}, {Carollo}, {Ilbert}, {Karim}, {Lefevre}, {Lilly},
  {Scoville}, {Smolcic}, \& {Yan}}]{Capak15}
{Capak}, P.~L., {Carilli}, C., {Jones}, G., {et~al.} 2015, \nat, 522, 455

\bibitem[{{Cappellaro} {et~al.}(2015){Cappellaro}, {Botticella}, {Pignata},
  {Grado}, {Greggio}, {Limatola}, {Vaccari}, {Baruffolo}, {Benetti}, {Bufano},
  {Capaccioli}, {Cascone}, {Covone}, {De Cicco}, {Falocco}, {Della Valle},
  {Jarvis}, {Marchetti}, {Napolitano}, {Paolillo}, {Pastorello}, {Radovich},
  {Schipani}, {Spiro}, {Tomasella}, \& {Turatto}}]{Cappellaro15}
{Cappellaro}, E., {Botticella}, M.~T., {Pignata}, G., {et~al.} 2015, \aap, 584,
  A62

\bibitem[{{Carilli} \& {Walter}(2013)}]{Carilli13}
{Carilli}, C.~L. \& {Walter}, F. 2013, \araa, 51, 105

\bibitem[{{Casavecchia} {et~al.}(2024){Casavecchia}, {Maio}, {P{\'e}roux}, \&
  {Ciardi}}]{Casavecchia24}
{Casavecchia}, B., {Maio}, U., {P{\'e}roux}, C., \& {Ciardi}, B. 2024, \aap,
  689, A106

\bibitem[{{Chabrier}(2003)}]{Chabrier03}
{Chabrier}, G. 2003, \pasp, 115, 763

\bibitem[{{CONCERTO Collaboration} {et~al.}(2020){CONCERTO Collaboration},
  {Ade}, {Aravena}, {Barria}, {Beelen}, {Benoit}, {B{\'e}thermin}, {Bounmy},
  {Bourrion}, {Bres}, {De Breuck}, {Calvo}, {Cao}, {Catalano}, {D{\'e}sert},
  {Dur{\'a}n}, {Fasano}, {Fenouillet}, {Garcia}, {Garde}, {Goupy}, {Groppi},
  {Hoarau}, {Lagache}, {Lambert}, {Leggeri}, {Levy-Bertrand},
  {Mac{\'\i}as-P{\'e}rez}, {Mani}, {Marpaud}, {Mauskopf}, {Monfardini},
  {Pisano}, {Ponthieu}, {Prieur}, {Roni}, {Roudier}, {Tourres}, \&
  {Tucker}}]{CONCERTO20}
{CONCERTO Collaboration}, {Ade}, P., {Aravena}, M., {et~al.} 2020, \aap, 642,
  A60

\bibitem[{{Contursi} {et~al.}(2017){Contursi}, {Baker}, {Berta}, {Magnelli},
  {Lutz}, {Fischer}, {Verma}, {Nielbock}, {Gr{\'a}cia Carpio}, {Veilleux},
  {Sturm}, {Davies}, {Genzel}, {Hailey-Dunsheath}, {Herrera-Camus}, {Janssen},
  {Poglitsch}, {Sternberg}, \& {Tacconi}}]{Contursi17}
{Contursi}, A., {Baker}, A.~J., {Berta}, S., {et~al.} 2017, \aap, 606, A86

\bibitem[{{Cormier} {et~al.}(2019){Cormier}, {Abel}, {Hony}, {Lebouteiller},
  {Madden}, {Polles}, {Galliano}, {De Looze}, {Galametz}, \&
  {Lambert-Huyghe}}]{Cormier19}
{Cormier}, D., {Abel}, N.~P., {Hony}, S., {et~al.} 2019, \aap, 626, A23

\bibitem[{{Cormier} {et~al.}(2015){Cormier}, {Madden}, {Lebouteiller}, {Abel},
  {Hony}, {Galliano}, {R{\'e}my-Ruyer}, {Bigiel}, {Baes}, {Boselli},
  {Chevance}, {Cooray}, {De Looze}, {Doublier}, {Galametz}, {Hughes},
  {Karczewski}, {Lee}, {Lu}, \& {Spinoglio}}]{Cormier15}
{Cormier}, D., {Madden}, S.~C., {Lebouteiller}, V., {et~al.} 2015, \aap, 578,
  A53

\bibitem[{{Croxall} {et~al.}(2017){Croxall}, {Smith}, {Pellegrini}, {Groves},
  {Bolatto}, {Herrera-Camus}, {Sandstrom}, {Draine}, {Wolfire}, {Armus},
  {Boquien}, {Brandl}, {Dale}, {Galametz}, {Hunt}, {Kennicutt}, {Kreckel},
  {Rigopoulou}, {van der Werf}, \& {Wilson}}]{Croxall17}
{Croxall}, K.~V., {Smith}, J.~D., {Pellegrini}, E., {et~al.} 2017, \apj, 845,
  96

\bibitem[{{Curtis-Lake} {et~al.}(2023){Curtis-Lake}, {Carniani}, {Cameron},
  {Charlot}, {Jakobsen}, {Maiolino}, {Bunker}, {Witstok}, {Smit}, {Chevallard},
  {Willott}, {Ferruit}, {Arribas}, {Bonaventura}, {Curti}, {D'Eugenio},
  {Franx}, {Giardino}, {Looser}, {L{\"u}tzgendorf}, {Maseda}, {Rawle}, {Rix},
  {Rodr{\'\i}guez del Pino}, {{\"U}bler}, {Sirianni}, {Dressler}, {Egami},
  {Eisenstein}, {Endsley}, {Hainline}, {Hausen}, {Johnson}, {Rieke},
  {Robertson}, {Shivaei}, {Stark}, {Tacchella}, {Williams}, {Willmer},
  {Bhatawdekar}, {Bowler}, {Boyett}, {Chen}, {de Graaff}, {Helton}, {Hviding},
  {Jones}, {Kumari}, {Lyu}, {Nelson}, {Perna}, {Sandles}, {Saxena}, {Suess},
  {Sun}, {Topping}, {Wallace}, \& {Whitler}}]{CurtisLake23}
{Curtis-Lake}, E., {Carniani}, S., {Cameron}, A., {et~al.} 2023, Nature
  Astronomy, 7, 622

\bibitem[{{Daddi} {et~al.}(2010){Daddi}, {Bournaud}, {Walter}, {Dannerbauer},
  {Carilli}, {Dickinson}, {Elbaz}, {Morrison}, {Riechers}, {Onodera}, {Salmi},
  {Krips}, \& {Stern}}]{Daddi10}
{Daddi}, E., {Bournaud}, F., {Walter}, F., {et~al.} 2010, \apj, 713, 686

\bibitem[{{Dav{\'e}} {et~al.}(2020){Dav{\'e}}, {Crain}, {Stevens}, {Narayanan},
  {Saintonge}, {Catinella}, \& {Cortese}}]{Dave20}
{Dav{\'e}}, R., {Crain}, R.~A., {Stevens}, A. R.~H., {et~al.} 2020, \mnras,
  497, 146

\bibitem[{{De Breuck} {et~al.}(2019){De Breuck}, {Wei{\ss}}, {B{\'e}thermin},
  {Cunningham}, {Apostolovski}, {Aravena}, {Archipley}, {Chapman}, {Chen},
  {Fu}, {Jarugula}, {Malkan}, {Mangian}, {Phadke}, {Reuter}, {Stacey},
  {Strandet}, {Vieira}, \& {Vishwas}}]{DeBreuck19}
{De Breuck}, C., {Wei{\ss}}, A., {B{\'e}thermin}, M., {et~al.} 2019, \aap, 631,
  A167

\bibitem[{{De Looze} {et~al.}(2014){De Looze}, {Cormier}, {Lebouteiller},
  {Madden}, {Baes}, {Bendo}, {Boquien}, {Boselli}, {Clements}, {Cortese},
  {Cooray}, {Galametz}, {Galliano}, {Graci{\'a}-Carpio}, {Isaak}, {Karczewski},
  {Parkin}, {Pellegrini}, {R{\'e}my-Ruyer}, {Spinoglio}, {Smith}, \&
  {Sturm}}]{DeLooze14}
{De Looze}, I., {Cormier}, D., {Lebouteiller}, V., {et~al.} 2014, \aap, 568,
  A62

\bibitem[{{Decarli} {et~al.}(2020){Decarli}, {Aravena}, {Boogaard}, {Carilli},
  {Gonz{\'a}lez-L{\'o}pez}, {Walter}, {Cortes}, {Cox}, {da Cunha}, {Daddi},
  {D{\'\i}az-Santos}, {Hodge}, {Inami}, {Neeleman}, {Novak}, {Oesch},
  {Popping}, {Riechers}, {Smail}, {Uzgil}, {van der Werf}, {Wagg}, \&
  {Weiss}}]{Decarli20}
{Decarli}, R., {Aravena}, M., {Boogaard}, L., {et~al.} 2020, \apj, 902, 110

\bibitem[{{Decarli} {et~al.}(2019){Decarli}, {Walter},
  {G{\'o}nzalez-L{\'o}pez}, {Aravena}, {Boogaard}, {Carilli}, {Cox}, {Daddi},
  {Popping}, {Riechers}, {Uzgil}, {Weiss}, {Assef}, {Bacon}, {Bauer},
  {Bertoldi}, {Bouwens}, {Contini}, {Cortes}, {da Cunha}, {D{\'\i}az-Santos},
  {Elbaz}, {Inami}, {Hodge}, {Ivison}, {Le F{\`e}vre}, {Magnelli}, {Novak},
  {Oesch}, {Rix}, {Sargent}, {Smail}, {Swinbank}, {Somerville}, {van der Werf},
  {Wagg}, \& {Wisotzki}}]{Decarli19}
{Decarli}, R., {Walter}, F., {G{\'o}nzalez-L{\'o}pez}, J., {et~al.} 2019, \apj,
  882, 138

\bibitem[{{Dessauges-Zavadsky} {et~al.}(2020){Dessauges-Zavadsky}, {Ginolfi},
  {Pozzi}, {B{\'e}thermin}, {Le F{\`e}vre}, {Fujimoto}, {Silverman}, {Jones},
  {Vallini}, {Schaerer}, {Faisst}, {Khusanova}, {Fudamoto}, {Cassata},
  {Loiacono}, {Capak}, {Yan}, {Amorin}, {Bardelli}, {Boquien}, {Cimatti},
  {Gruppioni}, {Hathi}, {Ibar}, {Koekemoer}, {Lemaux}, {Narayanan}, {Oesch},
  {Rodighiero}, {Romano}, {Talia}, {Toft}, {Vergani}, {Zamorani}, \&
  {Zucca}}]{Dessauges-Zavadsky20}
{Dessauges-Zavadsky}, M., {Ginolfi}, M., {Pozzi}, F., {et~al.} 2020, \aap, 643,
  A5

\bibitem[{{Di Cesare} {et~al.}(2024){Di Cesare}, {Ginolfi}, {Graziani},
  {Schneider}, {Romano}, \& {Popping}}]{DiCesare24}
{Di Cesare}, C., {Ginolfi}, M., {Graziani}, L., {et~al.} 2024, \aap, 690, A255

\bibitem[{{D{\'\i}az-Santos} {et~al.}(2017){D{\'\i}az-Santos}, {Armus},
  {Charmandaris}, {Lu}, {Stierwalt}, {Stacey}, {Malhotra}, {van der Werf},
  {Howell}, {Privon}, {Mazzarella}, {Goldsmith}, {Murphy}, {Barcos-Mu{\~n}oz},
  {Linden}, {Inami}, {Larson}, {Evans}, {Appleton}, {Iwasawa}, {Lord},
  {Sanders}, \& {Surace}}]{DiazSantos17}
{D{\'\i}az-Santos}, T., {Armus}, L., {Charmandaris}, V., {et~al.} 2017, \apj,
  846, 32

\bibitem[{{D{\'\i}az-Santos} {et~al.}(2013){D{\'\i}az-Santos}, {Armus},
  {Charmandaris}, {Stierwalt}, {Murphy}, {Haan}, {Inami}, {Malhotra},
  {Meijerink}, {Stacey}, {Petric}, {Evans}, {Veilleux}, {van der Werf}, {Lord},
  {Lu}, {Howell}, {Appleton}, {Mazzarella}, {Surace}, {Xu}, {Schulz},
  {Sanders}, {Bridge}, {Chan}, {Frayer}, {Iwasawa}, {Melbourne}, \&
  {Sturm}}]{DiazSantos13}
{D{\'\i}az-Santos}, T., {Armus}, L., {Charmandaris}, V., {et~al.} 2013, \apj,
  774, 68

\bibitem[{{Dobbs} \& {Pringle}(2013)}]{Dobbs13}
{Dobbs}, C.~L. \& {Pringle}, J.~E. 2013, \mnras, 432, 653

\bibitem[{{Draine} \& {Bertoldi}(1996)}]{Draine96}
{Draine}, B.~T. \& {Bertoldi}, F. 1996, \apj, 468, 269

\bibitem[{{Draine} \& {Sutin}(1987)}]{Draine87}
{Draine}, B.~T. \& {Sutin}, B. 1987, \apj, 320, 803

\bibitem[{{Dunne} {et~al.}(2022){Dunne}, {Maddox}, {Papadopoulos}, {Ivison}, \&
  {Gomez}}]{Dunne22}
{Dunne}, L., {Maddox}, S.~J., {Papadopoulos}, P.~P., {Ivison}, R.~J., \&
  {Gomez}, H.~L. 2022, \mnras, 517, 962

\bibitem[{{Ebagezio} {et~al.}(2023){Ebagezio}, {Seifried}, {Walch},
  {N{\"u}rnberger}, {Rathjen}, \& {Naab}}]{Ebagezio23}
{Ebagezio}, S., {Seifried}, D., {Walch}, S., {et~al.} 2023, \mnras, 525, 5631

\bibitem[{{Ferkinhoff} {et~al.}(2014){Ferkinhoff}, {Brisbin}, {Parshley},
  {Nikola}, {Stacey}, {Schoenwald}, {Higdon}, {Higdon}, {Verma}, {Riechers},
  {Hailey-Dunsheath}, {Menten}, {G{\"u}sten}, {Wei{\ss}}, {Irwin}, {Cho},
  {Niemack}, {Halpern}, {Amiri}, {Hasselfield}, {Wiebe}, {Ade}, \&
  {Tucker}}]{Ferkinhoff14}
{Ferkinhoff}, C., {Brisbin}, D., {Parshley}, S., {et~al.} 2014, \apj, 780, 142

\bibitem[{{Feruglio} {et~al.}(2023){Feruglio}, {Maio}, {Tripodi}, {Winters},
  {Zappacosta}, {Bischetti}, {Civano}, {Carniani}, {D'Odorico}, {Fiore},
  {Gallerani}, {Ginolfi}, {Maiolino}, {Piconcelli}, {Valiante}, \&
  {Zanchettin}}]{Feruglio23}
{Feruglio}, C., {Maio}, U., {Tripodi}, R., {et~al.} 2023, \apjl, 954, L10

\bibitem[{{Fukui} \& {Kawamura}(2010)}]{Fukui10}
{Fukui}, Y. \& {Kawamura}, A. 2010, \araa, 48, 547

\bibitem[{{Galli} \& {Palla}(1998)}]{Galli98}
{Galli}, D. \& {Palla}, F. 1998, \aap, 335, 403

\bibitem[{{Garcia} {et~al.}(2024){Garcia}, {Narayanan}, {Popping}, {Anirudh},
  {Sutherland}, \& {Kaasinen}}]{Garcia23}
{Garcia}, K., {Narayanan}, D., {Popping}, G., {et~al.} 2024, \apj, 974, 197

\bibitem[{{Gelli} {et~al.}(2024){Gelli}, {Mason}, \& {Hayward}}]{Gelli24}
{Gelli}, V., {Mason}, C., \& {Hayward}, C.~C. 2024, \apj, 975, 192

\bibitem[{{Genzel} {et~al.}(2015){Genzel}, {Tacconi}, {Lutz}, {Saintonge},
  {Berta}, {Magnelli}, {Combes}, {Garc{\'\i}a-Burillo}, {Neri}, {Bolatto},
  {Contini}, {Lilly}, {Boissier}, {Boone}, {Bouch{\'e}}, {Bournaud}, {Burkert},
  {Carollo}, {Colina}, {Cooper}, {Cox}, {Feruglio}, {F{\"o}rster Schreiber},
  {Freundlich}, {Gracia-Carpio}, {Juneau}, {Kovac}, {Lippa}, {Naab}, {Salome},
  {Renzini}, {Sternberg}, {Walter}, {Weiner}, {Weiss}, \& {Wuyts}}]{Genzel15}
{Genzel}, R., {Tacconi}, L.~J., {Lutz}, D., {et~al.} 2015, \apj, 800, 20

\bibitem[{{Gkogkou} {et~al.}(2023){Gkogkou}, {B{\'e}thermin}, {Lagache}, {Van
  Cuyck}, {Jullo}, {Aravena}, {Beelen}, {Benoit}, {Bounmy}, {Calvo},
  {Catalano}, {Cora}, {Croton}, {de la Torre}, {Fasano}, {Ferrara}, {Goupy},
  {Hoarau}, {Hu}, {Ishiyama}, {Knudsen}, {Lambert}, {Mac{\'\i}as-P{\'e}rez},
  {Marpaud}, {Mellema}, {Monfardini}, {Pallottini}, {Ponthieu}, {Prada},
  {Roehlly}, {Vallini}, \& {Walter}}]{Gkogkou23}
{Gkogkou}, A., {B{\'e}thermin}, M., {Lagache}, G., {et~al.} 2023, \aap, 670,
  A16

\bibitem[{{Glazer} {et~al.}(2024){Glazer}, {Brad{\u{a}}c}, {Sanders},
  {Fujimoto}, {Bolan}, {Ferrara}, {Strait}, {Jones}, {Lemaux}, {Vallini}, \&
  {Ryan}}]{Glazer24}
{Glazer}, K., {Brad{\u{a}}c}, M., {Sanders}, R.~L., {et~al.} 2024, \mnras, 531,
  945

\bibitem[{{Glover} \& {Clark}(2012)}]{Glover12}
{Glover}, S. C.~O. \& {Clark}, P.~C. 2012, \mnras, 421, 9

\bibitem[{{Gnedin} \& {Draine}(2014)}]{Gnedin14}
{Gnedin}, N.~Y. \& {Draine}, B.~T. 2014, \apj, 795, 37

\bibitem[{{Goldsmith} {et~al.}(2012){Goldsmith}, {Langer}, {Pineda}, \&
  {Velusamy}}]{Goldsmith12}
{Goldsmith}, P.~F., {Langer}, W.~D., {Pineda}, J.~L., \& {Velusamy}, T. 2012,
  \apjs, 203, 13

\bibitem[{{Greggio}(2005)}]{Greggio05}
{Greggio}, L. 2005, \aap, 441, 1055

\bibitem[{{Gullberg} {et~al.}(2015){Gullberg}, {De Breuck}, {Vieira},
  {Wei{\ss}}, {Aguirre}, {Aravena}, {B{\'e}thermin}, {Bradford}, {Bothwell},
  {Carlstrom}, {Chapman}, {Fassnacht}, {Gonzalez}, {Greve}, {Hezaveh},
  {Holzapfel}, {Husband}, {Ma}, {Malkan}, {Marrone}, {Menten}, {Murphy},
  {Reichardt}, {Spilker}, {Stark}, {Strandet}, \& {Welikala}}]{Gullberg15}
{Gullberg}, B., {De Breuck}, C., {Vieira}, J.~D., {et~al.} 2015, \mnras, 449,
  2883

\bibitem[{{Gurman} {et~al.}(2024){Gurman}, {Hu}, {Sternberg}, \& {van
  Dishoeck}}]{Gurman24}
{Gurman}, A., {Hu}, C.-Y., {Sternberg}, A., \& {van Dishoeck}, E.~F. 2024,
  \apj, 965, 179

\bibitem[{{Haardt} \& {Madau}(1996)}]{HaardtMadau96}
{Haardt}, F. \& {Madau}, P. 1996, \apj, 461, 20

\bibitem[{{Habing}(1968)}]{Habing68}
{Habing}, H.~J. 1968, \bain, 19, 421

\bibitem[{{Hamanowicz} {et~al.}(2023){Hamanowicz}, {Zwaan}, {P{\'e}roux},
  {Lagos}, {Klitsch}, {Ivison}, {Biggs}, {Szakacs}, \& {Fresco}}]{Hamanowicz23}
{Hamanowicz}, A., {Zwaan}, M.~A., {P{\'e}roux}, C., {et~al.} 2023, \mnras, 519,
  34

\bibitem[{{Hassan} \& {Gronke}(2021)}]{Hassan21}
{Hassan}, S. \& {Gronke}, M. 2021, \apj, 908, 219

\bibitem[{{Heintz} {et~al.}(2021){Heintz}, {Watson}, {Oesch}, {Narayanan}, \&
  {Madden}}]{Heintz21}
{Heintz}, K.~E., {Watson}, D., {Oesch}, P.~A., {Narayanan}, D., \& {Madden},
  S.~C. 2021, \apj, 922, 147

\bibitem[{{Hemmati} {et~al.}(2017){Hemmati}, {Yan}, {Diaz-Santos}, {Armus},
  {Capak}, {Faisst}, \& {Masters}}]{Hemmati17}
{Hemmati}, S., {Yan}, L., {Diaz-Santos}, T., {et~al.} 2017, \apj, 834, 36

\bibitem[{{Hernandez} {et~al.}(2023){Hernandez}, {Jones}, {Smith}, {Togi},
  {Aloisi}, {Blair}, {Hirschauer}, {Hunt}, {James}, {Kumari}, {Lebouteiller},
  {Mingozzi}, \& {Ramambason}}]{Hernandez23}
{Hernandez}, S., {Jones}, L., {Smith}, L.~J., {et~al.} 2023, \apj, 948, 124

\bibitem[{{Hollenbach} \& {McKee}(1979)}]{Hollenbach79}
{Hollenbach}, D. \& {McKee}, C.~F. 1979, \apjs, 41, 555

\bibitem[{{Hollenbach} \& {McKee}(1989)}]{Hollenbach89}
{Hollenbach}, D. \& {McKee}, C.~F. 1989, \apj, 342, 306

\bibitem[{{Hu} {et~al.}(2021){Hu}, {Sternberg}, \& {van Dishoeck}}]{Hu21}
{Hu}, C.-Y., {Sternberg}, A., \& {van Dishoeck}, E.~F. 2021, \apj, 920, 44

\bibitem[{{Hu} {et~al.}(2023){Hu}, {Sternberg}, \& {van Dishoeck}}]{Hu23}
{Hu}, C.-Y., {Sternberg}, A., \& {van Dishoeck}, E.~F. 2023, \apj, 952, 140

\bibitem[{{Hughes} {et~al.}(2017){Hughes}, {Ibar}, {Villanueva}, {Aravena},
  {Baes}, {Bourne}, {Cooray}, {Dunne}, {Dye}, {Eales}, {Furlanetto},
  {Herrera-Camus}, {Ivison}, {van Kampen}, {Lara-L{\'o}pez}, {Maddox},
  {Micha{\l}owski}, {Smith}, {Valiante}, {van der Werf}, \& {Xue}}]{Hughes17}
{Hughes}, T.~M., {Ibar}, E., {Villanueva}, V., {et~al.} 2017, \aap, 602, A49

\bibitem[{{Huynh} {et~al.}(2014){Huynh}, {Kimball}, {Norris}, {Smail}, {Chow},
  {Coppin}, {Emonts}, {Ivison}, {Smolcic}, \& {Swinbank}}]{Huynh14}
{Huynh}, M.~T., {Kimball}, A.~E., {Norris}, R.~P., {et~al.} 2014, \mnras, 443,
  L54

\bibitem[{{Ikeda} {et~al.}(2024){Ikeda}, {Tadaki}, {Mitsuhashi}, {Aravena}, {De
  Looze}, {F{\"o}rster Schreiber}, {Gonz{\'a}lez-L{\'o}pez}, {Herrera-Camus},
  {Spilker}, {Barcos-Mu{\~n}oz}, {da Cunha}, {Davies}, {D{\'\i}az-Santos},
  {Ferrara}, {Killi}, {Lee}, {Li}, {Lutz}, {Smit}, {Solimano}, {Telikova},
  {{\"U}bler}, {Veilleux}, \& {Villanueva}}]{Ikeda24}
{Ikeda}, R., {Tadaki}, K.-i., {Mitsuhashi}, I., {et~al.} 2024, arXiv e-prints,
  arXiv:2408.03374

\bibitem[{{Izotov} {et~al.}(2021){Izotov}, {Guseva}, {Fricke}, {Henkel},
  {Schaerer}, \& {Thuan}}]{Izotov21}
{Izotov}, Y.~I., {Guseva}, N.~G., {Fricke}, K.~J., {et~al.} 2021, \aap, 646,
  A138

\bibitem[{{Kaasinen} {et~al.}(2024){Kaasinen}, {Venemans}, {Harrington},
  {Boogaard}, {Meyer}, {Ba{\~n}ados}, {Decarli}, {Walter}, {Neeleman},
  {Rivera}, \& {da Cunha}}]{Kaasinen24}
{Kaasinen}, M., {Venemans}, B., {Harrington}, K.~C., {et~al.} 2024, \aap, 684,
  A33

\bibitem[{{Katz} {et~al.}(2019){Katz}, {Galligan}, {Kimm}, {Rosdahl},
  {Haehnelt}, {Blaizot}, {Devriendt}, {Slyz}, {Laporte}, \& {Ellis}}]{Katz19}
{Katz}, H., {Galligan}, T.~P., {Kimm}, T., {et~al.} 2019, \mnras, 487, 5902

\bibitem[{{Keating} {et~al.}(2020){Keating}, {Richings}, {Murray},
  {Faucher-Gigu{\`e}re}, {Hopkins}, {Wetzel}, {Kere{\v{s}}}, {Benincasa},
  {Feldmann}, {Loebman}, \& {Orr}}]{Keating20}
{Keating}, L.~C., {Richings}, A.~J., {Murray}, N., {et~al.} 2020, \mnras, 499,
  837

\bibitem[{{Kennicutt}(1998)}]{Kennicutt98}
{Kennicutt}, Robert~C., J. 1998, \apj, 498, 541

\bibitem[{{Khatri} {et~al.}(2024{\natexlab{a}}){Khatri}, {Porciani},
  {Romano-D{\'\i}az}, {Seifried}, \& {Sch{\"a}be}}]{Khatri24a}
{Khatri}, P., {Porciani}, C., {Romano-D{\'\i}az}, E., {Seifried}, D., \&
  {Sch{\"a}be}, A. 2024{\natexlab{a}}, \aap, 688, A194

\bibitem[{{Khatri} {et~al.}(2024{\natexlab{b}}){Khatri}, {Romano-D{\'\i}az}, \&
  {Porciani}}]{Khatri24b}
{Khatri}, P., {Romano-D{\'\i}az}, E., \& {Porciani}, C. 2024{\natexlab{b}},
  arXiv e-prints, arXiv:2411.09755

\bibitem[{{Klaassen} {et~al.}(2020){Klaassen}, {Mroczkowski}, {Cicone},
  {Hatziminaoglou}, {Sartori}, {De Breuck}, {Bryan}, {Dicker}, {Duran},
  {Groppi}, {Kaercher}, {Kawabe}, {Kohno}, \& {Geach}}]{Klaassen20}
{Klaassen}, P.~D., {Mroczkowski}, T.~K., {Cicone}, C., {et~al.} 2020, in
  Society of Photo-Optical Instrumentation Engineers (SPIE) Conference Series,
  Vol. 11445, Ground-based and Airborne Telescopes VIII, ed. H.~K. {Marshall},
  J.~{Spyromilio}, \& T.~{Usuda}, 114452F

\bibitem[{{Klitsch} {et~al.}(2019){Klitsch}, {P{\'e}roux}, {Zwaan}, {Smail},
  {Nelson}, {Popping}, {Chen}, {Diemer}, {Ivison}, {Allison}, {Muller},
  {Swinbank}, {Hamanowicz}, {Biggs}, \& {Dutta}}]{Klitsch19}
{Klitsch}, A., {P{\'e}roux}, C., {Zwaan}, M.~A., {et~al.} 2019, \mnras, 490,
  1220

\bibitem[{{Knudsen} {et~al.}(2016){Knudsen}, {Richard}, {Kneib}, {Jauzac},
  {Cl{\'e}ment}, {Drouart}, {Egami}, \& {Lindroos}}]{Knudsen16}
{Knudsen}, K.~K., {Richard}, J., {Kneib}, J.-P., {et~al.} 2016, \mnras, 462, L6

\bibitem[{{Kravtsov} \& {Belokurov}(2024)}]{Kravtsov24}
{Kravtsov}, A. \& {Belokurov}, V. 2024, arXiv e-prints, arXiv:2405.04578

\bibitem[{{Krumholz}(2013)}]{Krumholz13}
{Krumholz}, M.~R. 2013, \mnras, 436, 2747

\bibitem[{{Krumholz}(2014)}]{Krumholz14}
{Krumholz}, M.~R. 2014, \physrep, 539, 49

\bibitem[{{Krumholz} \& {Gnedin}(2011)}]{Krumholz11}
{Krumholz}, M.~R. \& {Gnedin}, N.~Y. 2011, \apj, 729, 36

\bibitem[{{Lagache} {et~al.}(2018){Lagache}, {Cousin}, \&
  {Chatzikos}}]{Lagache18}
{Lagache}, G., {Cousin}, M., \& {Chatzikos}, M. 2018, \aap, 609, A130

\bibitem[{{Lagos} {et~al.}(2015){Lagos}, {Crain}, {Schaye}, {Furlong}, {Frenk},
  {Bower}, {Schaller}, {Theuns}, {Trayford}, {Bah{\'e}}, \& {Dalla
  Vecchia}}]{Lagos15}
{Lagos}, C. d.~P., {Crain}, R.~A., {Schaye}, J., {et~al.} 2015, \mnras, 452,
  3815

\bibitem[{{Le F{\`e}vre} {et~al.}(2020){Le F{\`e}vre}, {B{\'e}thermin},
  {Faisst}, {Jones}, {Capak}, {Cassata}, {Silverman}, {Schaerer}, {Yan},
  {Amorin}, {Bardelli}, {Boquien}, {Cimatti}, {Dessauges-Zavadsky},
  {Giavalisco}, {Hathi}, {Fudamoto}, {Fujimoto}, {Ginolfi}, {Gruppioni},
  {Hemmati}, {Ibar}, {Koekemoer}, {Khusanova}, {Lagache}, {Lemaux}, {Loiacono},
  {Maiolino}, {Mancini}, {Narayanan}, {Morselli}, {M{\'e}ndez-Hern{\`a}ndez},
  {Oesch}, {Pozzi}, {Romano}, {Riechers}, {Scoville}, {Talia}, {Tasca},
  {Thomas}, {Toft}, {Vallini}, {Vergani}, {Walter}, {Zamorani}, \&
  {Zucca}}]{LeFevre20}
{Le F{\`e}vre}, O., {B{\'e}thermin}, M., {Faisst}, A., {et~al.} 2020, \aap,
  643, A1

\bibitem[{{Lepp} \& {Shull}(1983)}]{Lepp83}
{Lepp}, S. \& {Shull}, J.~M. 1983, \apj, 270, 578

\bibitem[{{Leroy} {et~al.}(2008){Leroy}, {Walter}, {Brinks}, {Bigiel}, {de
  Blok}, {Madore}, \& {Thornley}}]{Leroy08}
{Leroy}, A.~K., {Walter}, F., {Brinks}, E., {et~al.} 2008, \aj, 136, 2782

\bibitem[{{Li} {et~al.}(2024){Li}, {Da Cunha}, {Gonz{\'a}lez-L{\'o}pez},
  {Aravena}, {De Looze}, {F{\"o}rster Schreiber}, {Herrera-Camus}, {Spilker},
  {Tadaki}, {Barcos-Munoz}, {Battisti}, {Birkin}, {Bowler}, {Davies},
  {D{\'\i}az-Santos}, {Ferrara}, {Fisher}, {Hodge}, {Ikeda}, {Killi}, {Lee},
  {Liu}, {Lutz}, {Mitsuhashi}, {Naab}, {Posses}, {Rela{\~n}o}, {Solimano},
  {{\"U}bler}, {van der Giessen}, \& {Villanueva}}]{Juno24}
{Li}, J., {Da Cunha}, E., {Gonz{\'a}lez-L{\'o}pez}, J., {et~al.} 2024, \apj,
  976, 70

\bibitem[{{Liang} {et~al.}(2024){Liang}, {Feldmann}, {Murray}, {Narayanan},
  {Hayward}, {Angl{\'e}s-Alc{\'a}zar}, {Bassini}, {Richings},
  {Faucher-Gigu{\`e}re}, {Chung}, {Chan}, {Tolgay}, {{\c{C}}atmabacak},
  {Kere{\v{s}}}, \& {Hopkins}}]{Liang24}
{Liang}, L., {Feldmann}, R., {Murray}, N., {et~al.} 2024, \mnras, 528, 499

\bibitem[{{Lupi} \& {Bovino}(2020)}]{Lupi20}
{Lupi}, A. \& {Bovino}, S. 2020, \mnras, 492, 2818

\bibitem[{{Madden} {et~al.}(2020){Madden}, {Cormier}, {Hony}, {Lebouteiller},
  {Abel}, {Galametz}, {De Looze}, {Chevance}, {Polles}, {Lee}, {Galliano},
  {Lambert-Huyghe}, {Hu}, \& {Ramambason}}]{Madden20}
{Madden}, S.~C., {Cormier}, D., {Hony}, S., {et~al.} 2020, \aap, 643, A141

\bibitem[{{Magdis} {et~al.}(2014){Magdis}, {Rigopoulou}, {Hopwood}, {Huang},
  {Farrah}, {Pearson}, {Alonso-Herrero}, {Bock}, {Clements}, {Cooray},
  {Griffin}, {Oliver}, {Perez Fournon}, {Riechers}, {Swinyard}, {Scott},
  {Thatte}, {Valtchanov}, \& {Vaccari}}]{Magdis14}
{Magdis}, G.~E., {Rigopoulou}, D., {Hopwood}, R., {et~al.} 2014, \apj, 796, 63

\bibitem[{{Maio} {et~al.}(2019){Maio}, {Borgani}, {Ciardi}, \&
  {Petkova}}]{Maio19}
{Maio}, U., {Borgani}, S., {Ciardi}, B., \& {Petkova}, M. 2019, \pasa, 36, e020

\bibitem[{{Maio} {et~al.}(2010){Maio}, {Ciardi}, {Dolag}, {Tornatore}, \&
  {Khochfar}}]{Maio10}
{Maio}, U., {Ciardi}, B., {Dolag}, K., {Tornatore}, L., \& {Khochfar}, S. 2010,
  \mnras, 407, 1003

\bibitem[{{Maio} {et~al.}(2009){Maio}, {Ciardi}, {Yoshida}, {Dolag}, \&
  {Tornatore}}]{Maio09}
{Maio}, U., {Ciardi}, B., {Yoshida}, N., {Dolag}, K., \& {Tornatore}, L. 2009,
  \aap, 503, 25

\bibitem[{{Maio} {et~al.}(2007){Maio}, {Dolag}, {Ciardi}, \&
  {Tornatore}}]{Maio07}
{Maio}, U., {Dolag}, K., {Ciardi}, B., \& {Tornatore}, L. 2007, \mnras, 379,
  963

\bibitem[{{Maio} {et~al.}(2006){Maio}, {Dolag}, {Meneghetti}, {Moscardini},
  {Yoshida}, {Baccigalupi}, {Bartelmann}, \& {Perrotta}}]{Maio06}
{Maio}, U., {Dolag}, K., {Meneghetti}, M., {et~al.} 2006, \mnras, 373, 869

\bibitem[{{Maio} {et~al.}(2011{\natexlab{a}}){Maio}, {Khochfar}, {Johnson}, \&
  {Ciardi}}]{Maio11}
{Maio}, U., {Khochfar}, S., {Johnson}, J.~L., \& {Ciardi}, B.
  2011{\natexlab{a}}, \mnras, 414, 1145

\bibitem[{{Maio} {et~al.}(2011{\natexlab{b}}){Maio}, {Koopmans}, \&
  {Ciardi}}]{Maio11L}
{Maio}, U., {Koopmans}, L. V.~E., \& {Ciardi}, B. 2011{\natexlab{b}}, \mnras,
  412, L40

\bibitem[{{Maio} {et~al.}(2022){Maio}, {P{\'e}roux}, \& {Ciardi}}]{Maio22}
{Maio}, U., {P{\'e}roux}, C., \& {Ciardi}, B. 2022, \aap, 657, A47

\bibitem[{{Maio} {et~al.}(2016){Maio}, {Petkova}, {De Lucia}, \&
  {Borgani}}]{Maio16}
{Maio}, U., {Petkova}, M., {De Lucia}, G., \& {Borgani}, S. 2016, \mnras, 460,
  3733

\bibitem[{{Maio} \& {Viel}(2015)}]{Maio15}
{Maio}, U. \& {Viel}, M. 2015, \mnras, 446, 2760

\bibitem[{{Maio} \& {Viel}(2023)}]{Maio23}
{Maio}, U. \& {Viel}, M. 2023, \aap, 672, A71

\bibitem[{{Maoz} \& {Graur}(2017)}]{Maoz17}
{Maoz}, D. \& {Graur}, O. 2017, \apj, 848, 25

\bibitem[{{McMullin} {et~al.}(2007){McMullin}, {Waters}, {Schiebel}, {Young},
  \& {Golap}}]{McMullin07}
{McMullin}, J.~P., {Waters}, B., {Schiebel}, D., {Young}, W., \& {Golap}, K.
  2007, in Astronomical Society of the Pacific Conference Series, Vol. 376,
  Astronomical Data Analysis Software and Systems XVI, ed. R.~A. {Shaw},
  F.~{Hill}, \& D.~J. {Bell}, 127

\bibitem[{{Mitsuhashi} {et~al.}(2024){Mitsuhashi}, {Tadaki}, {Ikeda},
  {Herrera-Camus}, {Aravena}, {De Looze}, {F{\"o}rster Schreiber},
  {Gonz{\'a}lez-L{\'o}pez}, {Spilker}, {Assef}, {Bouwens}, {Barcos-Munoz},
  {Birkin}, {Bowler}, {Rivera}, {Davies}, {Da Cunha}, {D{\'\i}az-Santos},
  {Ferrara}, {Fisher}, {Lee}, {Li}, {Lutz}, {Rela{\~n}o}, {Naab}, {Palla},
  {Posses}, {Solimano}, {Tacconi}, {{\"U}bler}, {van der Giessen}, \&
  {Veilleux}}]{Mitsuhashi24}
{Mitsuhashi}, I., {Tadaki}, K.-i., {Ikeda}, R., {et~al.} 2024, \aap, 690, A197

\bibitem[{{Mu{\~n}oz-Elgueta} {et~al.}(2024){Mu{\~n}oz-Elgueta}, {Arrigoni
  Battaia}, {Kauffmann}, {Pakmor}, {Walch}, {Obreja}, \& {Buhlmann}}]{Munoz24}
{Mu{\~n}oz-Elgueta}, N., {Arrigoni Battaia}, F., {Kauffmann}, G., {et~al.}
  2024, \aap, 690, A392

\bibitem[{{Nagamine} {et~al.}(2004){Nagamine}, {Springel}, \&
  {Hernquist}}]{Nagamine04}
{Nagamine}, K., {Springel}, V., \& {Hernquist}, L. 2004, \mnras, 348, 421

\bibitem[{{Olsen} {et~al.}(2017){Olsen}, {Greve}, {Narayanan}, {Thompson},
  {Dav{\'e}}, {Niebla Rios}, \& {Stawinski}}]{Olsen17}
{Olsen}, K., {Greve}, T.~R., {Narayanan}, D., {et~al.} 2017, \apj, 846, 105

\bibitem[{{Olsen} {et~al.}(2018){Olsen}, {Greve}, {Narayanan}, {Thompson},
  {Dav{\'e}}, {Niebla Rios}, \& {Stawinski}}]{Olsen18}
{Olsen}, K., {Greve}, T.~R., {Narayanan}, D., {et~al.} 2018, \apj, 857, 148

\bibitem[{{Pallottini} {et~al.}(2017){Pallottini}, {Ferrara}, {Bovino},
  {Vallini}, {Gallerani}, {Maiolino}, \& {Salvadori}}]{Pallottini17}
{Pallottini}, A., {Ferrara}, A., {Bovino}, S., {et~al.} 2017, \mnras, 471, 4128

\bibitem[{{Pallottini} {et~al.}(2019){Pallottini}, {Ferrara}, {Decataldo},
  {Gallerani}, {Vallini}, {Carniani}, {Behrens}, {Kohandel}, \&
  {Salvadori}}]{Pallottini19}
{Pallottini}, A., {Ferrara}, A., {Decataldo}, D., {et~al.} 2019, \mnras, 487,
  1689

\bibitem[{{Patrick} {et~al.}(2023){Patrick}, {Whalen}, {Latif}, \&
  {Elford}}]{Patrick23}
{Patrick}, S.~J., {Whalen}, D.~J., {Latif}, M.~A., \& {Elford}, J.~S. 2023,
  \mnras, 522, 3795

\bibitem[{{P{\'e}roux} \& {Howk}(2020)}]{Peroux20}
{P{\'e}roux}, C. \& {Howk}, J.~C. 2020, \araa, 58, 363

\bibitem[{{P{\'e}roux} {et~al.}(2019){P{\'e}roux}, {Zwaan}, {Klitsch},
  {Augustin}, {Hamanowicz}, {Rahmani}, {Pettini}, {Kulkarni}, {Straka},
  {Biggs}, {York}, \& {Milliard}}]{Peroux19}
{P{\'e}roux}, C., {Zwaan}, M.~A., {Klitsch}, A., {et~al.} 2019, \mnras, 485,
  1595

\bibitem[{{Pineda} {et~al.}(2013){Pineda}, {Langer}, {Velusamy}, \&
  {Goldsmith}}]{Pineda13}
{Pineda}, J.~L., {Langer}, W.~D., {Velusamy}, T., \& {Goldsmith}, P.~F. 2013,
  \aap, 554, A103

\bibitem[{{Popping} {et~al.}(2019){Popping}, {Narayanan}, {Somerville},
  {Faisst}, \& {Krumholz}}]{Popping19}
{Popping}, G., {Narayanan}, D., {Somerville}, R.~S., {Faisst}, A.~L., \&
  {Krumholz}, M.~R. 2019, \mnras, 482, 4906

\bibitem[{{Popping} \& {P{\'e}roux}(2022)}]{Popping22}
{Popping}, G. \& {P{\'e}roux}, C. 2022, \mnras, 513, 1531

\bibitem[{{Ramos Padilla} {et~al.}(2021){Ramos Padilla}, {Wang}, {Ploeckinger},
  {van der Tak}, \& {Trager}}]{RamosPadilla21}
{Ramos Padilla}, A.~F., {Wang}, L., {Ploeckinger}, S., {van der Tak}, F.~F.~S.,
  \& {Trager}, S.~C. 2021, \aap, 645, A133

\bibitem[{{Ramos Padilla} {et~al.}(2023){Ramos Padilla}, {Wang}, {van der Tak},
  \& {Trager}}]{RamosPadilla23}
{Ramos Padilla}, A.~F., {Wang}, L., {van der Tak}, F.~F.~S., \& {Trager}, S.~C.
  2023, \aap, 679, A131

\bibitem[{{R{\'e}my-Ruyer} {et~al.}(2014){R{\'e}my-Ruyer}, {Madden},
  {Galliano}, {Galametz}, {Takeuchi}, {Asano}, {Zhukovska}, {Lebouteiller},
  {Cormier}, {Jones}, {Bocchio}, {Baes}, {Bendo}, {Boquien}, {Boselli},
  {DeLooze}, {Doublier-Pritchard}, {Hughes}, {Karczewski}, \&
  {Spinoglio}}]{Remy-Ruyer14}
{R{\'e}my-Ruyer}, A., {Madden}, S.~C., {Galliano}, F., {et~al.} 2014, \aap,
  563, A31

\bibitem[{{Riechers} {et~al.}(2019){Riechers}, {Pavesi}, {Sharon}, {Hodge},
  {Decarli}, {Walter}, {Carilli}, {Aravena}, {da Cunha}, {Daddi}, {Dickinson},
  {Smail}, {Capak}, {Ivison}, {Sargent}, {Scoville}, \& {Wagg}}]{Riechers19}
{Riechers}, D.~A., {Pavesi}, R., {Sharon}, C.~E., {et~al.} 2019, \apj, 872, 7

\bibitem[{{Righi} {et~al.}(2008){Righi}, {Hern{\'a}ndez-Monteagudo}, \&
  {Sunyaev}}]{Righi08}
{Righi}, M., {Hern{\'a}ndez-Monteagudo}, C., \& {Sunyaev}, R.~A. 2008, \aap,
  489, 489

\bibitem[{{Roussel} {et~al.}(2007){Roussel}, {Helou}, {Hollenbach}, {Draine},
  {Smith}, {Armus}, {Schinnerer}, {Walter}, {Engelbracht}, {Thornley},
  {Kennicutt}, {Calzetti}, {Dale}, {Murphy}, \& {Bot}}]{Roussel07}
{Roussel}, H., {Helou}, G., {Hollenbach}, D.~J., {et~al.} 2007, \apj, 669, 959

\bibitem[{{Roy} \& {Lapi}(2024)}]{Roy24}
{Roy}, A. \& {Lapi}, A. 2024, arXiv e-prints, arXiv:2407.19007

\bibitem[{{Saintonge} {et~al.}(2017){Saintonge}, {Catinella}, {Tacconi},
  {Kauffmann}, {Genzel}, {Cortese}, {Dav{\'e}}, {Fletcher},
  {Graci{\'a}-Carpio}, {Kramer}, {Heckman}, {Janowiecki}, {Lutz}, {Rosario},
  {Schiminovich}, {Schuster}, {Wang}, {Wuyts}, {Borthakur}, {Lamperti}, \&
  {Roberts-Borsani}}]{Saintonge17}
{Saintonge}, A., {Catinella}, B., {Tacconi}, L.~J., {et~al.} 2017, \apjs, 233,
  22

\bibitem[{{Saintonge} {et~al.}(2011){Saintonge}, {Kauffmann}, {Kramer},
  {Tacconi}, {Buchbender}, {Catinella}, {Fabello}, {Graci{\'a}-Carpio}, {Wang},
  {Cortese}, {Fu}, {Genzel}, {Giovanelli}, {Guo}, {Haynes}, {Heckman},
  {Krumholz}, {Lemonias}, {Li}, {Moran}, {Rodriguez-Fernandez}, {Schiminovich},
  {Schuster}, \& {Sievers}}]{Saintonge11}
{Saintonge}, A., {Kauffmann}, G., {Kramer}, C., {et~al.} 2011, \mnras, 415, 32

\bibitem[{{Salpeter}(1955)}]{Salpeter55}
{Salpeter}, E.~E. 1955, \apj, 121, 161

\bibitem[{{Salvestrini} {et~al.}(2024){Salvestrini}, {Feruglio}, {Tripodi},
  {Fontanot}, {Bischetti}, {De Lucia}, {Fiore}, {Hirschmann}, {Maio},
  {Piconcelli}, {Saccheo}, {Tortosa}, {Valiante}, {Xie}, \&
  {Zappacosta}}]{Salvestrini24}
{Salvestrini}, F., {Feruglio}, C., {Tripodi}, R., {et~al.} 2024, arXiv
  e-prints, arXiv:2412.02688

\bibitem[{{Saslaw} \& {Zipoy}(1967)}]{Saslaw67}
{Saslaw}, W.~C. \& {Zipoy}, D. 1967, \nat, 216, 976

\bibitem[{{Schaerer} {et~al.}(2015){Schaerer}, {Boone}, {Jones},
  {Dessauges-Zavadsky}, {Sklias}, {Zamojski}, {Cava}, {Richard}, {Ellis},
  {Rawle}, {Egami}, \& {Combes}}]{Schaerer15}
{Schaerer}, D., {Boone}, F., {Jones}, T., {et~al.} 2015, \aap, 576, L2

\bibitem[{{Schaerer} {et~al.}(2020){Schaerer}, {Ginolfi}, {B{\'e}thermin},
  {Fudamoto}, {Oesch}, {Le F{\`e}vre}, {Faisst}, {Capak}, {Cassata},
  {Silverman}, {Yan}, {Jones}, {Amorin}, {Bardelli}, {Boquien}, {Cimatti},
  {Dessauges-Zavadsky}, {Giavalisco}, {Hathi}, {Fujimoto}, {Ibar}, {Koekemoer},
  {Lagache}, {Lemaux}, {Loiacono}, {Maiolino}, {Narayanan}, {Morselli},
  {M{\'e}ndez-Hern{\`a}ndez}, {Pozzi}, {Riechers}, {Talia}, {Toft}, {Vallini},
  {Vergani}, {Zamorani}, \& {Zucca}}]{Schaerer20}
{Schaerer}, D., {Ginolfi}, M., {B{\'e}thermin}, M., {et~al.} 2020, \aap, 643,
  A3

\bibitem[{{Schaerer} {et~al.}(2022){Schaerer}, {Marques-Chaves}, {Barrufet},
  {Oesch}, {Izotov}, {Naidu}, {Guseva}, \& {Brammer}}]{Schaerer22}
{Schaerer}, D., {Marques-Chaves}, R., {Barrufet}, L., {et~al.} 2022, \aap, 665,
  L4

\bibitem[{{Schimek} {et~al.}(2024){Schimek}, {Decataldo}, {Shen}, {Cicone},
  {Baumschlager}, {van Kampen}, {Klaassen}, {Madau}, {Di Mascolo}, {Mayer},
  {Montoya Arroyave}, {Mroczkowski}, \& {Warraich}}]{Schimek24}
{Schimek}, A., {Decataldo}, D., {Shen}, S., {et~al.} 2024, \aap, 682, A98

\bibitem[{{Sommovigo} {et~al.}(2021){Sommovigo}, {Ferrara}, {Carniani},
  {Zanella}, {Pallottini}, {Gallerani}, \& {Vallini}}]{Sommovigo2021}
{Sommovigo}, L., {Ferrara}, A., {Carniani}, S., {et~al.} 2021, \mnras, 503,
  4878

\bibitem[{{Springel}(2005)}]{Springel05}
{Springel}, V. 2005, \mnras, 364, 1105

\bibitem[{{Springel} \& {Hernquist}(2003)}]{Springel03}
{Springel}, V. \& {Hernquist}, L. 2003, \mnras, 339, 289

\bibitem[{{Stacey} {et~al.}(1991){Stacey}, {Geis}, {Genzel}, {Lugten},
  {Poglitsch}, {Sternberg}, \& {Townes}}]{Stacey91}
{Stacey}, G.~J., {Geis}, N., {Genzel}, R., {et~al.} 1991, \apj, 373, 423

\bibitem[{{Szakacs} {et~al.}(2021){Szakacs}, {P{\'e}roux}, {Zwaan},
  {Hamanowicz}, {Klitsch}, {Fresco}, {Augustin}, {Biggs}, {Kulkarni}, \&
  {Rahmani}}]{Szakacs21}
{Szakacs}, R., {P{\'e}roux}, C., {Zwaan}, M., {et~al.} 2021, \mnras, 505, 4746

\bibitem[{{Szakacs} {et~al.}(2022){Szakacs}, {P{\'e}roux}, {Zwaan}, {Nelson},
  {Schinnerer}, {Lah{\'e}n}, {Weng}, \& {Fresco}}]{Szakacs22}
{Szakacs}, R., {P{\'e}roux}, C., {Zwaan}, M.~A., {et~al.} 2022, \mnras, 512,
  4736

\bibitem[{{Tacconi} {et~al.}(2018){Tacconi}, {Genzel}, {Saintonge}, {Combes},
  {Garc{\'\i}a-Burillo}, {Neri}, {Bolatto}, {Contini}, {F{\"o}rster Schreiber},
  {Lilly}, {Lutz}, {Wuyts}, {Accurso}, {Boissier}, {Boone}, {Bouch{\'e}},
  {Bournaud}, {Burkert}, {Carollo}, {Cooper}, {Cox}, {Feruglio}, {Freundlich},
  {Herrera-Camus}, {Juneau}, {Lippa}, {Naab}, {Renzini}, {Salome}, {Sternberg},
  {Tadaki}, {{\"U}bler}, {Walter}, {Weiner}, \& {Weiss}}]{Tacconi18}
{Tacconi}, L.~J., {Genzel}, R., {Saintonge}, A., {et~al.} 2018, \apj, 853, 179

\bibitem[{{Tacconi} {et~al.}(2020){Tacconi}, {Genzel}, \&
  {Sternberg}}]{Tacconi20}
{Tacconi}, L.~J., {Genzel}, R., \& {Sternberg}, A. 2020, \araa, 58, 157

\bibitem[{{Tacconi} {et~al.}(2013){Tacconi}, {Neri}, {Genzel}, {Combes},
  {Bolatto}, {Cooper}, {Wuyts}, {Bournaud}, {Burkert}, {Comerford}, {Cox},
  {Davis}, {F{\"o}rster Schreiber}, {Garc{\'\i}a-Burillo}, {Gracia-Carpio},
  {Lutz}, {Naab}, {Newman}, {Omont}, {Saintonge}, {Shapiro Griffin}, {Shapley},
  {Sternberg}, \& {Weiner}}]{Tacconi13}
{Tacconi}, L.~J., {Neri}, R., {Genzel}, R., {et~al.} 2013, \apj, 768, 74

\bibitem[{{Tarantino} {et~al.}(2021){Tarantino}, {Bolatto}, {Herrera-Camus},
  {Harris}, {Wolfire}, {Buchbender}, {Croxall}, {Dale}, {Groves}, {Levy},
  {Riquelme}, {Smith}, \& {Stutzki}}]{Tarantino21}
{Tarantino}, E., {Bolatto}, A.~D., {Herrera-Camus}, R., {et~al.} 2021, \apj,
  915, 92

\bibitem[{{Telford} {et~al.}(2023){Telford}, {McQuinn}, {Chisholm}, \&
  {Berg}}]{Telford23}
{Telford}, O.~G., {McQuinn}, K. B.~W., {Chisholm}, J., \& {Berg}, D.~A. 2023,
  \apj, 943, 65

\bibitem[{{Telikova} {et~al.}(2024){Telikova}, {Gonz{\'a}lez-L{\'o}pez},
  {Aravena}, {Posses}, {Villanueva}, {Baeza-Garay}, {Jones}, {Solimano}, {Lee},
  {Assef}, {De Looze}, {Diaz Santos}, {Ferrara}, {Ikeda}, {Herrera-Camus},
  {{\"U}bler}, {Lamperti}, {Mitsuhashi}, {Relano}, {Perna}, \&
  {Tadaki}}]{Telikova24}
{Telikova}, K., {Gonz{\'a}lez-L{\'o}pez}, J., {Aravena}, M., {et~al.} 2024,
  arXiv e-prints, arXiv:2411.09033

\bibitem[{{Tinsley}(1980)}]{Tinsley80}
{Tinsley}, B.~M. 1980, \fcp, 5, 287

\bibitem[{{Togi} \& {Smith}(2016)}]{Togi16}
{Togi}, A. \& {Smith}, J.~D.~T. 2016, \apj, 830, 18

\bibitem[{{Tornatore} {et~al.}(2007){Tornatore}, {Borgani}, {Dolag}, \&
  {Matteucci}}]{Tornatore07}
{Tornatore}, L., {Borgani}, S., {Dolag}, K., \& {Matteucci}, F. 2007, \mnras,
  382, 1050

\bibitem[{{Vallini} {et~al.}(2015){Vallini}, {Gallerani}, {Ferrara},
  {Pallottini}, \& {Yue}}]{Vallini15}
{Vallini}, L., {Gallerani}, S., {Ferrara}, A., {Pallottini}, A., \& {Yue}, B.
  2015, \apj, 813, 36

\bibitem[{{Vizgan} {et~al.}(2022{\natexlab{a}}){Vizgan}, {Greve}, {Olsen},
  {Zanella}, {Narayanan}, {Dav{\`e}}, {Magdis}, {Popping}, {Valentino}, \&
  {Heintz}}]{Vizgan22H2}
{Vizgan}, D., {Greve}, T.~R., {Olsen}, K.~P., {et~al.} 2022{\natexlab{a}},
  \apj, 929, 92

\bibitem[{{Vizgan} {et~al.}(2022{\natexlab{b}}){Vizgan}, {Heintz}, {Greve},
  {Narayanan}, {Dav{\'e}}, {Olsen}, {Popping}, \& {Watson}}]{Vizgan22HI}
{Vizgan}, D., {Heintz}, K.~E., {Greve}, T.~R., {et~al.} 2022{\natexlab{b}},
  \apjl, 939, L1

\bibitem[{{Walter} {et~al.}(2022){Walter}, {Neeleman}, {Decarli}, {Venemans},
  {Meyer}, {Weiss}, {Ba{\~n}ados}, {Bosman}, {Carilli}, {Fan}, {Riechers},
  {Rix}, \& {Thompson}}]{Walter22}
{Walter}, F., {Neeleman}, M., {Decarli}, R., {et~al.} 2022, \apj, 927, 21

\bibitem[{{Weibel} {et~al.}(2024){Weibel}, {de Graaff}, {Setton}, {Miller},
  {Oesch}, {Brammer}, {Lagos}, {Whitaker}, {Williams}, {Baggen}, {Bezanson},
  {Boogaard}, {Cleri}, {Greene}, {Hirschmann}, {Hviding}, {Kuruvanthodi},
  {Labb{\'e}}, {Leja}, {Maseda}, {Matthee}, {McConachie}, {Naidu},
  {Roberts-Borsani}, {Schaerer}, {Suess}, {Valentino}, {van Dokkum}, \&
  {Wang}}]{Weibel24}
{Weibel}, A., {de Graaff}, A., {Setton}, D.~J., {et~al.} 2024, arXiv e-prints,
  arXiv:2409.03829

\bibitem[{{Williams} {et~al.}(2024){Williams}, {Lake}, {Naoz}, {Burkhart},
  {Treu}, {Marinacci}, {Nakazato}, {Vogelsberger}, {Yoshida}, {Chiaki},
  {Chiou}, \& {Chen}}]{Williams24}
{Williams}, C.~E., {Lake}, W., {Naoz}, S., {et~al.} 2024, \apjl, 960, L16

\bibitem[{{Witten} {et~al.}(2024){Witten}, {McClymont}, {Laporte},
  {Roberts-Borsani}, {Sijacki}, {Tacchella}, {Simmonds}, {Katz}, {Ellis},
  {Witstok}, {Maiolino}, {Ji}, {Hayes}, {Looser}, \& {D'Eugenio}}]{Witten24}
{Witten}, C., {McClymont}, W., {Laporte}, N., {et~al.} 2024, arXiv e-prints,
  arXiv:2407.07937

\bibitem[{{Woods} {et~al.}(2024){Woods}, {Patrick}, {Whalen}, \&
  {Heger}}]{Woods24}
{Woods}, T.~E., {Patrick}, S., {Whalen}, D.~J., \& {Heger}, A. 2024, \apj, 960,
  59

\bibitem[{{Yamaguchi} {et~al.}(2017){Yamaguchi}, {Kohno}, {Tamura}, {Oguri},
  {Ezawa}, {Hayatsu}, {Kitayama}, {Matsuda}, {Matsuo}, {Oshima}, {Ota},
  {Izumi}, \& {Umehata}}]{Yamaguchi17}
{Yamaguchi}, Y., {Kohno}, K., {Tamura}, Y., {et~al.} 2017, \apj, 845, 108

\bibitem[{{Yan} {et~al.}(2020){Yan}, {Sajina}, {Loiacono}, {Lagache},
  {B{\'e}thermin}, {Faisst}, {Ginolfi}, {F{\`e}vre}, {Gruppioni}, {Capak},
  {Cassata}, {Schaerer}, {Silverman}, {Bardelli}, {Dessauges-Zavadsky},
  {Cimatti}, {Hathi}, {Lemaux}, {Ibar}, {Jones}, {Koekemoer}, {Oesch}, {Talia},
  {Pozzi}, {Riechers}, {Tasca}, {Toft}, {Vallini}, {Vergani}, {Zamorani}, \&
  {Zucca}}]{Yan20}
{Yan}, L., {Sajina}, A., {Loiacono}, F., {et~al.} 2020, \apj, 905, 147

\bibitem[{{Yang} {et~al.}(2022){Yang}, {Popping}, {Somerville}, {Pullen},
  {Breysse}, \& {Maniyar}}]{Yang22}
{Yang}, S., {Popping}, G., {Somerville}, R.~S., {et~al.} 2022, \apj, 929, 140

\bibitem[{{Yoshida} {et~al.}(2003){Yoshida}, {Sokasian}, {Hernquist}, \&
  {Springel}}]{Yoshida03}
{Yoshida}, N., {Sokasian}, A., {Hernquist}, L., \& {Springel}, V. 2003, \apjl,
  591, L1

\bibitem[{{Zanella} {et~al.}(2018){Zanella}, {Daddi}, {Magdis}, {Diaz Santos},
  {Cormier}, {Liu}, {Cibinel}, {Gobat}, {Dickinson}, {Sargent}, {Popping},
  {Madden}, {Bethermin}, {Hughes}, {Valentino}, {Rujopakarn}, {Pannella},
  {Bournaud}, {Walter}, {Wang}, {Elbaz}, \& {Coogan}}]{Zanella18}
{Zanella}, A., {Daddi}, E., {Magdis}, G., {et~al.} 2018, \mnras, 481, 1976

\end{thebibliography}


\begin{appendix}

\section{\CII\ 158~$\mu$m line emission data}
\label{sect:appendix}

Here we provide the atomic data adopted in this work and in the companion paper C24 to compute the \CII\ 158~$\mu$m line emission.
We model the fine-structure transition $(2p)[\rm {}^2 P_{3/2} - {}^2 P_{1/2}]$ between the quantum number $J = 3/2$ and $J = 1/2$ of C$^+$ as a two-level system considering $e^-$ impacts, H impacts and H$_2$ impacts. 
Collisional rates, spontaneous-transition rate, and energy level separation are taken from \cite{Hollenbach89}, \cite{Maio07} and \cite{Goldsmith12} and listed below:
\begin{equation*}
    \gamma ^{ { \rm e}}_{21} = 2.8\cdot 10^{-7} \, { T_{100}^{-0.5}}\; \; \; { \rm cm^3 \: s^{-1}},
\end{equation*}
\begin{equation*}
    \gamma ^{ { \rm H}}_{21} = 8\cdot 10^{-10} \, { T_{100}^{0.07}}\; \; \; {\rm cm^3 \: s^{-1}},
\end{equation*}
\begin{equation*}
    \gamma ^{ { \rm H_2}}_{21} = 3.8\cdot 10^{-10} \, { T_{100}^{0.14}}\; \; \; {\rm cm^3 \: s^{-1}},
\end{equation*}
\begin{equation*}
    A_{21} = 2.4 \cdot 10^{-6 } \; { \rm s^{-1}},
\end{equation*}
\begin{equation*}
    \Delta E_{21} = 1.259 \cdot 10^{-14 } \; {\rm  erg},
\end{equation*}
\noindent
where the indeces $2$ and $1$ refer to the transition upper and lower levels, $T_{100} = T/(100\,\rm K)$ with $T$ gas temperature, and $\Delta E_{21}$ corresponds to the transition wavelength (frequency) of 157.74~$\mu$m (1900.5~GHz), while the excitation temperature is about 91.2~K.
With these inputs, the \CII\ radiative losses can be computed as:
\begin{equation}
\Lambda_{\textup{\CII}} =  
\frac{ \left( n_{\rm H_2}\gamma_{12}^{\rm H_2} + n_{\rm H}\gamma_{12}^{\rm H} + n_{\textup{\rm e}}\gamma_{12}^{\rm e} \right) \, 
{n}_{{\textup{C}^+}} \, A_{21} \, \Delta E_{21}
}
{
{ n_{\rm H_2} \left(\gamma_{12}^{\rm H_2} + \gamma_{21}^{\rm H_2} \right) + n_{\textup{\rm H}} \left(\gamma_{12}^{\rm H} + \gamma_{21}^{\rm H} \right) + n_{\textup{\rm e}} \left(\gamma_{12}^{\rm e} + \gamma_{21}^{\rm e} \right) } + A_{21}
},
\label{eq:LambdaExtended}
\end{equation}
which highligths the contributions of H$_2$, \HI\ and $e^-$ (ionised/\HII) gas phases for given H$_2$, \HI, $e^-$ and C$^+$ number densities
($ n_{\rm H_2} $,
$ n_{\rm H} $,
$ n_{\rm e} $,
$ n_{\rm C^+} $, respectively).
The relation between excitation ($1\rightarrow2$ transition) and de-excitation ($2\rightarrow1$ transition) rates satisfies the Boltzmann limit \cite[see e.g.][]{Maio07}.
The \CII\ luminosity emitted by a gas volume $V$ can be estimated 
as \LCII~$=\Lambda_{\textup{\CII}}{V}$.

\end{appendix}


\end{document}